\documentclass{aa}

\usepackage[english]{babel}

\usepackage{natbib}
\usepackage{amsmath}
\usepackage{graphicx}
\usepackage{xcolor}
\usepackage[colorlinks=true, allcolors=blue]{hyperref}

\title{Modeling the UV-photon irradiation of CS$_2$-bearing ices in the laboratory with the {\sl pyRate} gas-grain astrochemical code}
\subtitle{New insights into the missing sulfur problem}
\titlerunning{Modeling ice laboratory experiments with {\sl pyRate}} 
\author{O. Sipil\"a\inst{1}, R. Mart\'in-Dom\'enech\inst{2}, W. Riedel\inst{1}, D. Navarro-Almaida\inst{2}, A. Fuente\inst{2}, A. Taillard\inst{2}, G.M. Muñoz Caro\inst{2}
}
\authorrunning{O. Sipil\"a\inst{1}, R. Mart\'in-Dom\'enech\inst{2}, et al.}
\institute{Max-Planck-Institut f\"ur Extraterrestrische Physik (MPE), Giessenbachstr. 1, 85748 Garching, Germany \\
e-mail: \texttt{osipila@mpe.mpg.de}
\and{Centro de Astrobiolog\'ia (CSIC-INTA) Carretera de Ajalvir, km. 4, Torrej\'on de Ardoz, E-28850 Madrid, Spain}
}

\begin{document}

\abstract{Observations indicate that the total abundance of S-bearing species in dense clouds is orders of magnitude lower than the cosmic sulfur abundance. Addressing this ``missing sulfur problem'' requires a combination of astronomical observations, laboratory experiments, and theoretical models. In this work, we use the {\sl pyRate} astrochemical model to simulate the VUV photon irradiation of a CO$_2$:CS$_2$ ice mixture at 10 K in the laboratory, with the goal of supporting the interpretation of the experimental results and testing our current understanding of the sulfur evolution in interstellar ices. For this purpose, the astrochemical model was adapted to the experimental conditions, and the chemical network was compiled from several sources to ensure that all known reactions involving sulfur species were included. The results indicate that nondiffusive chemistry is necessary to reproduce the formation of S-bearing species observed in the experiment. However, some discrepancies were found in the major S-bearing ice chemistry products predicted by the model and the experiment. The compounds OCS, CS, and SO are overpredicted by the model, while it falls short in accounting for $\rm SO_2$ and sulfur allotropes. These discrepancies are likely due to a combination of an incomplete knowledge of the chemical reactions at play (either because of missing reactions and/or because of unconstrained reaction barriers), and uncertainties in the experimental analysis. This work represents the first effort to model the chemistry of a multicomponent ice analog with a rate-equation based code, and highlights the complementary nature of theoretical and experimental astrochemistry to disentangle the chemical evolution of sulfur in the interstellar medium.}

\maketitle

\section{Introduction}

The chemistry of sulfur (S) in the interstellar medium (ISM) is not fully constrained. 
Observed abundances of S-bearing molecules in dense regions of the ISM are up to two orders of magnitude lower than the cosmic sulfur abundance \citep[see, e.g.,][]{vastel18,legal19,fuente19,riviere20}, with some variability depending on the environmental conditions \citep{fuente23}.  
This issue, historically known as the sulfur depletion problem or the missing sulfur problem, needs to be addressed through a combination of astronomical observations, laboratory experiments, and theoretical models. 

A significant fraction of the sulfur present in dense and cold interstellar regions is expected to be locked in icy dust grains \citep[see, e.g.,][]{laas19}. 
Sulfur molecules in interstellar ice mantles would be subject to energetic processing through cosmic ray or ultraviolet (UV) photon irradiation. These processes can be further investigated in the laboratory under conditions relevant to the ISM
to better understand the chemical evolution of sulfur in interstellar ices.
The S-bearing molecules more often studied in experiments simulating the energetic processing of ice analogs are hydrogen sulfide \citep[H$_2$S, see, e.g.,][]{antonio11,antonio12,antonio14,loeffler15,loeffler16,loeffler18,cazaux22,julia24,hector24} and sulfur dioxide \citep[SO$_2$, see, e.g.,][]{ferrante08,garozzo08,garozzo10,kanuchova17,nguyen24}. 
This is due to their relatively high abundances compared to other S-bearing species in comets and other bodies of the Solar System \citep[][]{dalton10,leroy15,calmonte16}. 
More recently, additional S-bearing molecules such as CS$_2$ have also been incorporated into such experimental studies, based on their detection in comets and their potential presence in  interstellar ices.

\citet{MartinDomenech24} presented laboratory experiments studying the chemistry induced by energetic processing of CS$_2$-bearing  
ices. 
In these experiments, binary ice samples containing one isotopically-labeled major ice component (H$_2^{18}$O, $^{13}$C$^{18}$O, or $^{13}$C$^{18}$O$_2$) and CS$_2$ molecules were co-deposited onto a cryogenically cooled substrate (down to a temperature of 7$-$11 K) located at the center of an ultra-high-vacuum chamber (with a base pressure of the order of $\sim$2 $\times$ 10$^{-10}$ Torr). 
The purpose of using isotopically-labeled molecules was to facilitate the subsequent analysis of the experimental results.
After deposition, the ice samples were irradiated with either 2 keV electrons 
or vacuum-ultraviolet (VUV) photons, 
while being monitored with infrared (IR) spectroscopy. 
After irradiation, the irradiated ices were warmed up to 250 K at a controlled heating rate of 2 K min$^{-1}$, and the desorbing molecules were detected with a
quadrupole mass spectrometer (QMS). 
Formation of a variety of S-bearing products was reported, including SO$_2$, OCS, SO$_3$, C$_3$S$_2$, and S$_2$ (and/or their isotopically labeled counterparts). 
However, a significant fraction of the initial amount of sulfur was not detected at the end of the experiments. 
The authors suggested that the missing sulfur could be contained in long sulfur allotropes (S$_n$, with $n\ge4$). 
These species have been proposed in the literature as potential carriers of the missing sulfur in dense regions of the ISM \citep[see, e.g.,][]{cazaux22,hector24}. 
Their plausible presence in the irradiated ices was compatible with the non-detection of a fraction of the initial sulfur, because sulfur allotropes do not have strong IR features, and those with $n\ge4$ have molecular masses above the upper limit of the mass-to-charge ratio detectable by the QMS in the experimental setup (100 amu). Moreover, the expected desorption temperature of S$_8$ (the most stable sulfur allotrope on Earth) is higher than 250 K \citep{perrero24}. As a result, their detection during the reported experiments was very challenging. This stressed the need for further experimental and/or theoretical work that could support the proposed scenario. 

Gas-grain astrochemical models are used to theoretically calculate the chemical composition over time of the solid and gaseous phases in different regions of the ISM. 
To this end, rate-equation based or stochastic Monte Carlo codes are used, in combination with a chemical network and a physical model.
Rate-equation based codes are usually more convenient to describe macroscopic processes. 
Stochastic codes, on the other hand, are able to provide more detailed information about microscopic processes that take place on surfaces, at the expense of greater computational cost \citep{cuppen17,izaskun25}. 
Because of their more accurate description of surface processes, stochastic codes have also been used to reproduce the ice chemistry observed in laboratory experiments. 
In particular, \citet{lamberts13,lamberts14} and \citet{ioppolo21} used the Continuous-Time Random-Walk Kinetic Monte Carlo (CTRW-KMC) astrochemical model \citep{cuppen07}, with a restricted chemical network, to model the formation of water and glycine (respectively) through nonenergetic processing of ice analogs in the laboratory. However, the application of these models to the simulation of more complex laboratory experiments is not necessarily practical due to the exponentially higher computational cost associated with such systems. 
Alternatively, \citet{Shingledecker19} used the \texttt{MONACO} rate-equation based astrochemical model \citep{vasyunin17} to reproduce the experimental results of pure O$_2$ and H$_2$O ice bombardment with keV protons. 
More recently, \citet{sokolova26} have compared the outcomes of the rate-equation based astrochemical models \texttt{MONACO}, \texttt{Nautilus} \citep{ruaud16}, and {\sl pyRate} \citep{Sipila15a} reproducing the hydrogenation of pure CO ices.
Their findings suggest that rate-equation based astrochemical models may also be suitable for reproducing laboratory experiments and facilitating their analysis. 
In this regard, \citet{pilling22} developed a rate-equation based code called \texttt{PROCODA} specifically designed to characterize the chemistry of irradiated astrophysical ice analogs  by calculating reaction rates using the ice column densities obtained from absorbance infrared (IR) spectra. 
This approach has been applied to the irradiation of CO$_2$ \citep{pilling22}, H$_2$O \citep{dasilveira24}, and CH$_4$ \citep{gerasimenko25} ices, as well as a H$_2$O:O$_2$ ice mixture \citep{silva25}. 

In this work, we use the astrochemical model {\sl pyRate} to simulate the laboratory experiments presented in \cite{MartinDomenech24}. 
The goal is twofold: to test our current understanding of the primary chemical reactions governing the evolution of S-bearing species in interstellar ices, and to provide predictions for species that may have formed during the experiments but could not be detected.
This work represents the first effort to model the chemistry of a multicomponent ice analog with a rate-equation based astrochemical model. This paper is structured as follows. In Sect.\,\ref{s:model}, we discuss the details of the simulations, describing how the experimental setup can be mimicked with a rate-equation chemical code and how the chemical simulation is initialized. Section~\ref{s:results} presents the results of our simulations, which are discussed in further detail in Sect.\,\ref{s:discussion}. The astrophysical implications of our work are discussed in Sect.\,\ref{s:implications}, and our final conclusions are drawn in Sect.\,\ref{s:conclusions}. Appendices~\ref{aa:activeLayers}~to~\ref{aa:photorates} present supplementary results as well as the chemical network used in this work.

\section{Model}\label{s:model}

Chemical simulations were carried out using our gas-grain astrochemical code {\sl pyRate}, whose basic functionality is described in \citet{Sipila15a}, with updates and added functionality presented in subsequent works \citep[e.g.,][]{Sipila19a}. 
The simulation setup is presented in Sect. \ref{sec:sim}. 
In \citet{MartinDomenech24}, multiple experiments with varying ice thicknesses and initial compositions were performed. The ice samples were irradiated with either 2\,keV electrons or VUV photons. 
Because molecular dissociation in {\sl pyRate} occurs mostly through photoprocesses (cosmic rays can also act as a dissociating agent) and there is no existing implementation for electron-induced dissociation, we chose to simulate the VUV photon irradiation of a CO$_2$\footnote{As mentioned in the Introduction, the experiments presented in \citet{MartinDomenech24} used isotopes ($^{13}$C$^{18}$O$_2$) in order to facilitate the analysis of the IR and QMS data. Since this was not required for the simulations presented in this work, the regular isotopolog CO$_2$ has been used instead.}:CS$_2$ ice mixture, more specifically their experiment~5. The adaptation of the astrochemical model to this particular ice sample is described in Sect. \ref{sec:desc}.

\subsection{Simulation setup}\label{sec:sim}
Briefly, {\sl pyRate} is a rate-equation based chemical code that solves a system of coupled ordinary differential equations. 
Even though {\sl pyRate} was developed to simulate chemistry occurring in interstellar environments, it is adaptable to other contexts, and has been used to simulate an ion trap experiment \citep{JimenezRedondo24}. Here, we simulate instead a different type of experiment where layers of ice have been grown on a substrate, and chemical reactions are initiated after irradiation of the ice breaks up the initially deposited molecules. We have used {\sl pyRate} to mimic the experiment by setting up the simulation such that, initially, the simulated ``grains'' are covered with ice corresponding to the amount grown on the substrate for the experiment. The surface area of one ``grain'' is set to correspond to the surface area of the substrate in the experiment. For simplicity, we tuned the various simulation parameters (e.g., gas density, which is used to calculate the grain number density) such that there was exactly one ``grain'' per cubic centimeter, which allowed to easily calculate the initial amount of ice to be deposited on the virtual substrate. The ice was then subjected to a UV field which started the chemical evolution along the reaction pathways involved in the chemical network (see Sect. \ref{sec:react}), and the simulation was run for the same amount of time as the duration of the experiment.
 
In a typical {\sl pyRate} simulation, the chemical evolution is resolved simultaneously in the gas phase and on grain surfaces; the two are linked via adsorption and desorption. 
There are four possible mechanisms for desorption: thermal, cosmic-ray induced, reactive (i.e., chemical), and photodesorption. 
As the experiments involve a low substrate temperature and there are no weakly-bound species such as hydrogen, thermal desorption over the timescale of the experiment is almost negligible (but was included in the simulation nonetheless). 
Cosmic-ray induced desorption was turned off as we do not expect it to be of any significance considering the short timescale of the experiment and the low flux of cosmic rays making it to the surface of the Earth. 
Reactive desorption was implemented under the assumption of 1\% desorption efficiency for exothermic reactions involving one or two products \citep{Riedel23}. 
Photodesorption was not included in the simulations as the photodesorption yields for most of the species involved in the present reaction scheme are not known.  
We did however test the effect of including photodesorption for $\rm CO_2$ and $\rm CS_2$ based on the flux of VUV photons in the experiment ($5.3 \times 10^{13} \, \rm photons \, cm^{-2} \, s^{-1}$; \citealt{MartinDomenech24}). For the photodesorption yields we assumed $10^{-3}$ for $\rm CO_2$ \citep{Oberg09b}, and (arbitrarily) the same value for $\rm CS_2$. We found that the photodesorption rates were too small to have an appreciable effect on ice abundances over the course of the simulation, and hence in all simulations described below photodesorption was not included for simplicity. 
Typically, the total amount of ice remains constant within one per cent over the timescale of the simulation. Although transfer from the gas phase back onto the ice is possible via adsorption, we have found this to be of a similarly negligible effect over the course of the experiment. Indeed the evolution of the ice is almost completely driven by in situ reactivity.
Finally, we employed the so-called three-phase model where the ice is separated into an active surface and a chemically inert mantle beneath (following \citealt{Hasegawa93b}). The motivation for this and the associated assumptions are further described below.

\subsection{Description of molecular dissociation}\label{sec:desc}

As explained above, 
we chose to simulate
experiment~5 in \citet{MartinDomenech24} (see their Table~1). Therefore, we set the initial ice composition to 530 monolayers (MLs) in total, divided between $\rm CO_2$ and $\rm CS_2$ in a 93:7 ratio. The temperature of the ice was set to 10\,K. To facilitate the simulation of this ice sample, two practical problems need to be addressed: the fraction of chemically active ice, and the photodissociation behavior of the ice molecules.

Firstly, the amount of energy absorbed in the ice over the timescale of the experiment could not be determined (even though the flux of the VUV photons in the experimental setup is known), because ices do not absorb the energy of the incoming photons at 100\% efficiency and the UV absorption cross section of $\rm CS_2$ is not known (see also \citealt{MartinDomenech24}). 
This means that the fraction of the ice that was ``activated'' by the incoming UV photons is not known either. 
Furthermore, the action of the UV flux on the ice is not constant in the sense that the bottom part of the ice experiences an attenuated UV flux compared to the surface due to the absorption of the top ice layers, following the Beer-Lambert law \citep[see, e.g.,][]{gustavo14}. 
We mimic these processes by adopting the three-phase chemical model where only some fraction of the ice layers (the so-called ice surface) are chemically active. The caveat that the appropriate number of active layers -- that is, the relative amount of the ice that is affected by the UV flux -- is unconstrained in the experiment is managed by testing a series of models with varying thickness for the active surface; we explored values ranging from one~ML to 300~MLs. Based on the test results we chose 100 active MLs as the fiducial case. This choice is further motivated in Appendix~\ref{aa:activeLayers} where the results of the test models are described.

\begin{figure}
\centering
        \includegraphics[width=.8\columnwidth]{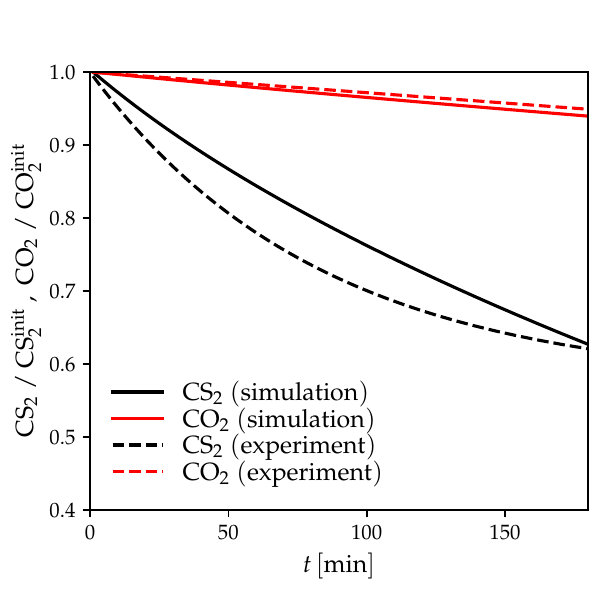}
    \caption{Time-dependent decay of $\rm CO_2$ (red lines) and $\rm CS_2$ (black lines), normalized to the initial amount of $\rm CO_2$ and $\rm CS_2$, over the course of the experiment (dashed lines). Solid lines show the corresponding data as predicted by our fiducial model (see text).}
        \label{fig:CS2_CO2_decay}
\end{figure}

Secondly, it is not straightforward to set up the description of dissociation in the simulation so that it matches the experimental conditions -- the simulation needs to be calibrated to the experimental setup.
In the chemical simulations the rate coefficients of the photoprocesses (for the most part taken from the KIDA network \citep{Wakelam15}; see also Appendix\,\ref{aa:chemicalNetwork}) do not directly depend on the (V)UV flux, but are instead parameterized expressions that depend on the visual extinction, determined based on the interstellar UV flux described in \citet{Draine78}\footnote{See also the documentation on the KIDA webpage: https://kida.astrochem-tools.org}. The efficiency of photodissociation in the simulation can be tuned by scaling the strength of the Draine field (the scaling factor is termed the ``$G_0$ factor'') as well as the visual extinction. For simplicity, we decided to not modify the $\rm CO_2$ and $\rm CS_2$ photodissociation rate coefficients, and searched for the best match between the simulations and experiments by setting the visual extinction to zero and testing different values of $G_0$ to obtain dissociation curves for $\rm CO_2$ and $\rm CS_2$ that resemble those derived from the experimental data -- the best-fit model was considered to be the one which minimizes the difference in the amounts of $\rm CO_2$ and $\rm CS_2$ between the simulations and the experiment after 180 minutes\footnote{This corresponds to the duration of the VUV irradiation in the experiments of \citet{MartinDomenech24}. All simulations presented in this paper have been run over a simulation time of 180\,min.}. Figure~\ref{fig:CS2_CO2_decay} shows these data for the fiducial model (see also below), corresponding to $G_0 = 6 \times 10^3$. The decay of $\rm CO_2$ is predicted extremely well by the simulation, and also for $\rm CS_2$ the dissociation curve is very similar to the experimental one even if the slope of the exponential differs somewhat between the two. We emphasize, however, that our present results do not imply that in interstellar conditions the ice is expected to be processed to this extent by incoming UV radiation. Indeed, it is noted in \citet{MartinDomenech24} that the UV fluence in their experiments is approximately an order of magnitude higher than the average UV fluence experienced by interstellar ices over a period of $2 \times 10^6 \, \rm yr$ in typical molecular cloud conditions.

\subsection{Chemical network; description of reactivity}\label{sec:react}

Given that we are simulating an ice experiment, we have not considered any gas-phase reactions. The chemical network for the ice chemistry -- containing only molecules involving the elements O, C, and S -- has been compiled from several sources and is reproduced in Appendix~\ref{aa:chemicalNetwork}.

Gas-grain chemical models like {\sl pyRate} customarily treat grain-surface chemistry under the assumption that reaction rates are governed by the diffusion of the reactants on or in the ice (see \citealt{Sipila15a} for the implementation in {\sl pyRate}), and usually the diffusion is assumed to occur thermally. However, the low mobility of heavy and strongly-bound species at low temperatures ($\sim10 \, \rm K$) inhibits the formation of large molecules, and it has been suggested that nondiffusive processes are required to explain the observed presence of complex organic molecules in the gas phase (resulting from grain-surface formation followed by desorption) in cold clouds \citep{Shingledecker19,Jin20,Garrod22}. The idea is that a product of a chemical reaction taking place in the ice can immediately react with a molecule in a neighboring binding site without the need for diffusion, greatly boosting the formation rates of large molecules even at low temperatures. The initiating reaction can occur in several ways: diffusively via the Langmuir-Hinshelwood mechanism\footnote{In this work we only consider thermal diffusion. In pyRate diffusion via tunneling is also an option, but applied to hydrogen only, which the present reaction system does not include.}, via the Eley-Rideal mechanism where an incoming atom from the gas phase directly reacts with a surface molecule, or via photodissociation of a surface molecule. We include all three possibilities in the present model, though since we are simulating ice chemistry only and re-adsorption of any species is a very minor channel, the Eley-Rideal mechanism plays a very marginal role in the present setup. However, we do expect nondiffusive reactions to be very important in the present case because the experiments involve only species that are relatively immobile in the ice. Hence we have run simulations with diffusive and nondiffusive chemistry, with the latter implemented following the numerical prescriptions of \citet{Jin20} and \citet{Garrod22}. Further details on the nondiffusive chemistry implementation in {\sl pyRate} are given in \citet{Riedel25}.

\section{Results}\label{s:results}

\subsection{Diffusive chemistry}

\begin{figure*}
\sidecaption
\centering
        \includegraphics[width=12cm]{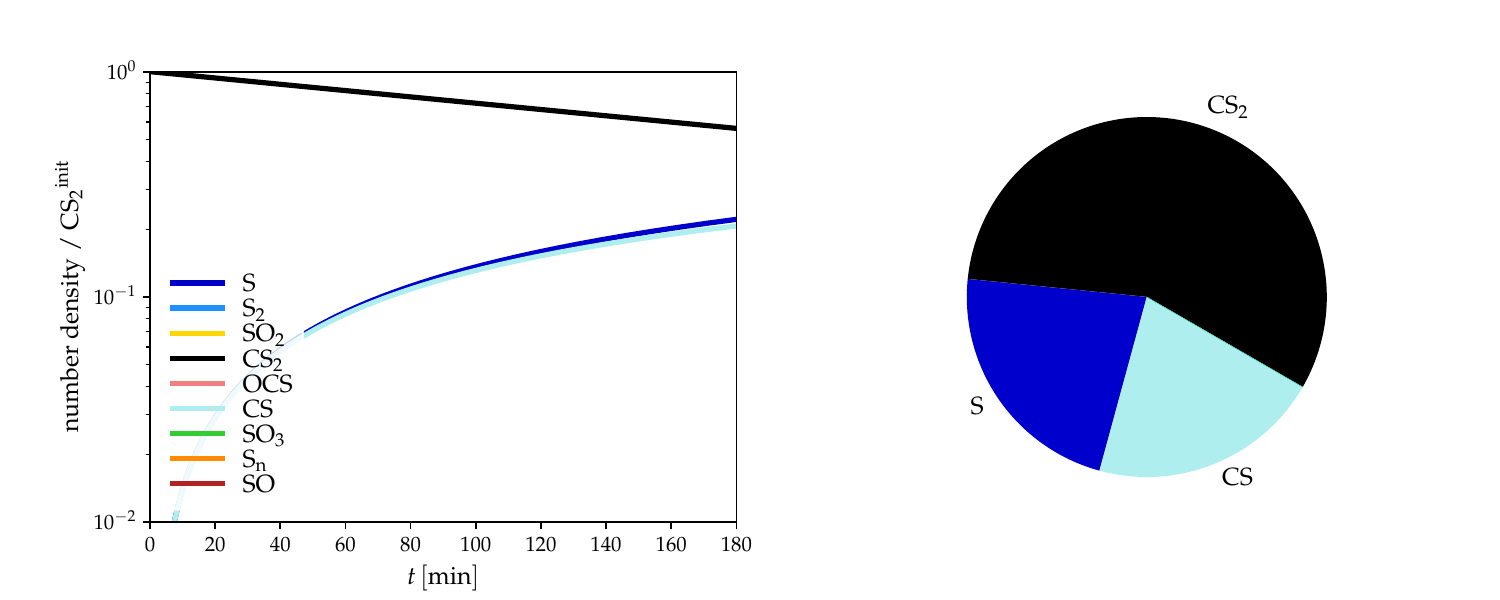}
    \caption{Left: Time-evolution of the number densities of selected sulfur carriers, normalized to the initial number density of $\rm CS_2$, in the simulation corresponding to a 100\,ML thick active ice surface layer. Right: pie chart showing the distribution of S at the end of the simulation. In both panels, the number densities of each species have been multiplied by the atomic sulfur count (i.e., taking into account that there are two S atoms in $\rm CS_2$, etc.), meaning that the figures display the distribution of the elemental sulfur reservoir across the species. Here, the shorthand ${\rm S}_n \equiv \sum_{n>2} {\rm S}_n$.
    }
        \label{fig:diff100ML}
\end{figure*}

Figure~\ref{fig:diff100ML} shows the results of the fiducial model when only diffusive chemistry is considered; here we concentrate on sulfur-containing species. Evidently, almost no reactivity occurs over the course of the simulation -- the only species that are produced in appreciable amounts are S and CS, which are the direct products of $\rm CS_2$ dissociation (of course O and CO form via $\rm CO_2$ dissociation as well, though not shown on the Figure). Other species (as indicated in the legend) that are expected to be produced via reactions between the dissociation products are formed at number densities approximately ten orders of magnitude below those of S and CS. There is no significant difference in the results if the thickness of the active surface layer is increased. The results hence show clearly that the scenario where reactivity is limited by the diffusion of the reactants is inadequate for describing the results of the experiments, which involve heavy species that are largely immobile at low temperatures. This suggests that diffusive chemistry alone is not appropriate to describe ice chemistry in cold dense cores with $T< 10$\,K in the UV-shielded interior regions, where sulfur-containing (complex) molecules are formed. This is especially important for the chemistry involving heavy atoms such as sulfur and those formed through reactions of heavy molecules such as CO$_2$.

\subsection{Nondiffusive chemistry}\label{sec:nondiff}

\begin{figure*}
\centering
\begin{picture}(600,150)(0,0)
\put(0,0){
\begin{picture}(0,0) 
        \includegraphics[width=1.5\columnwidth]{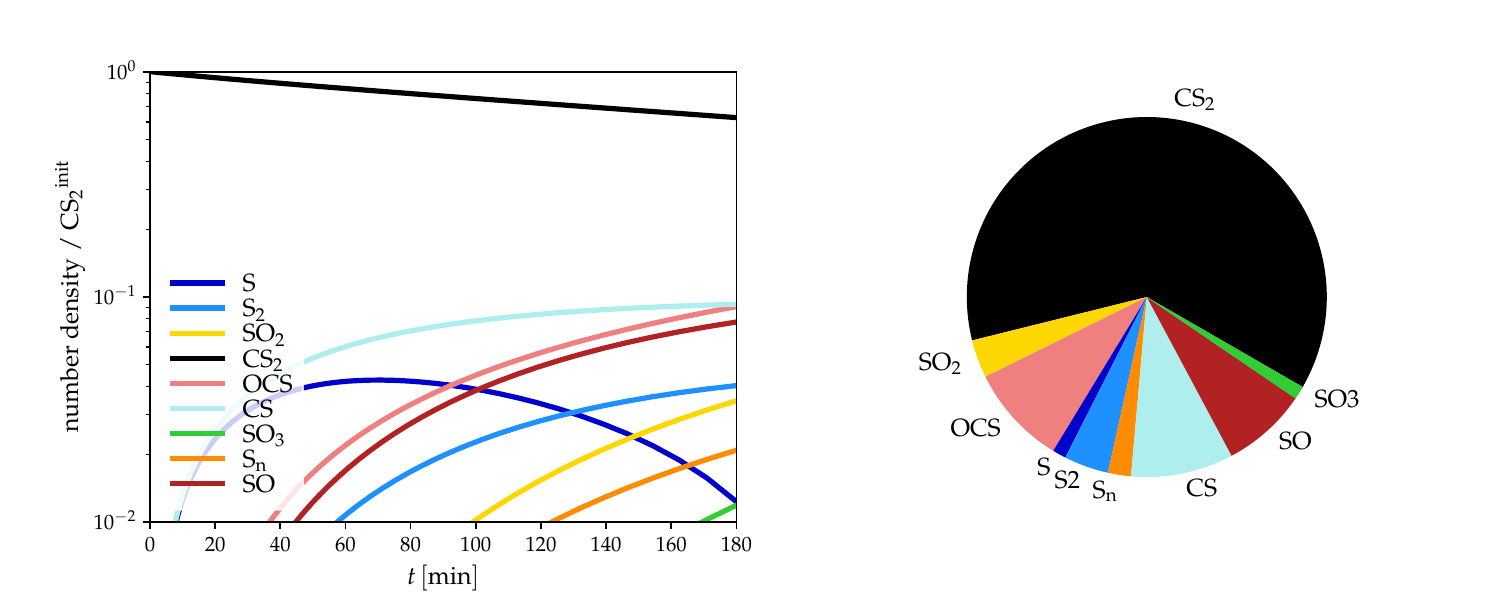}
\end{picture}}
\put(370,22){
\begin{picture}(0,0) 
        \includegraphics[width=0.49\columnwidth]{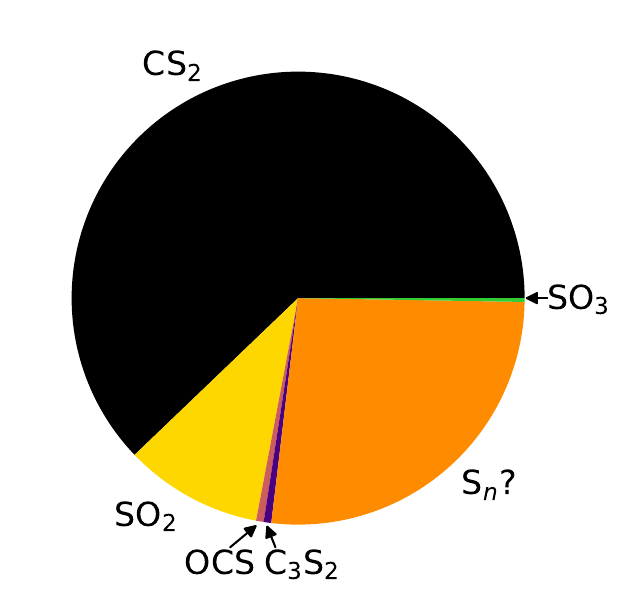}
\end{picture}}
\end{picture}  
    \caption{Left and middle: As Fig.\,\ref{fig:diff100ML}, but showing the results of the fiducial simulation applying nondiffusive chemistry. Right: distribution of sulfur at the end of experiment~5 in \citet{MartinDomenech24}.}
        \label{fig:nondiff100ML}
\end{figure*}

To boost the formation of the various sulfur-bearing species whose presence is expected based on the experiments, nondiffusive chemistry is invoked. Figure~\ref{fig:nondiff100ML} displays the results of a model otherwise identical to that shown in Fig.\,\ref{fig:diff100ML}, but with nondiffusive chemistry enabled, along with the results of the experiment.
To ease the interpretation of Fig.\,\ref{fig:nondiff100ML} and all simulation figures that follow, the product molecules have been grouped according to oxygen content, with species with 0/1/2/3 oxygen atoms shown in shade of blue/red/yellow/green, respectively. ${\rm S}_n$ is separately identified in orange as a special case.
In this case, reactivity proceeds to a much larger extent than in the scenario with only diffusive chemistry, closer to what was observed in the experiments. Yet, a comparison of the simulation results to the experimental ones reveals many interesting differences between the two. 

The simulation predicts high number densities of OCS, CS, and SO, while in the experiments only a small amount of OCS is formed and detection of CS and SO was not reported. These three species are chemically connected. CS is a direct product of $\rm CS_2$ dissociation, which also forms atomic S. 
OCS forms via the association of either CS and O, or CO and S (with CO and O originating in the dissociation of $\rm CO_2$). 
SO is in turn produced via $\rm O + S \longrightarrow SO$ or $\rm OCS + O \longrightarrow CO + SO$.
Given the rather elementary reactions involved in the reaction network, it is a puzzling contradiction that these species are not detected in the experiment (with the exception of small amounts of OCS) while they are produced efficiently in the simulation. In fact, as it can be seen from the time-evolution displayed in Fig.\,\ref{fig:nondiff100ML}, OCS is, throughout the simulation, the most abundant molecule that is not a direct product of $\rm CS_2$ or $\rm CO_2$ dissociation. 
The formation pathways of OCS, CS, and SO$_2$ in the model are further evaluated in Sect. \ref{ss:uncertainties}. 

Another major difference between the results of the simulations and the experiment is in the number density of $\rm SO_2$ and ${\rm S}_n$. 
In the experiment, $\rm SO_2$ is produced efficiently (see also the figures in \citealt{MartinDomenech24}) while a significant fraction ($\sim$25 \%) of atomic sulfur was suggested to be in the form of sulfur allotropes.
The simulation conversely struggles somewhat with the production of $\rm SO_2$ and only a small fraction of atomic S ends up in allotropes. This discrepancy may be due to the fact that a significant portion of atomic S is in the simulations contained in OCS, CS, and SO, which does not appear to be the case in the experiment. In addition, excitation of atomic S produced upon photodissociation of CS$_2$ molecules (which is currently not taken into account in {\sl pyRate}) could play a role in the formation of allotropes. This possibility is explored in Sect. \ref{ss:excitation}.

Our results strongly suggest that the chemical network governing the production and destruction of the involved species is not known in sufficient detail. In Sections \ref{ss:uncertainties} and \ref{ss:excitation}, we discuss potential modifications to the chemical network, as well as other features of the model, that may lead to a better match between the simulation and experimental results. 
In addition, Section \ref{sec:as_so_exp} revisits the analysis of the experiment to explore the non detection of CS and SO, and to evaluate other uncertainties that could also contribute to the discrepancies found between the simulation and the experimental results.
Finally, we note that in the case of nondiffusive chemistry, the number densities recovered at the end of the simulation do depend on the assumed thickness of the active ice surface layer. As explained in Sect. \ref{sec:desc}, we quantify this in Appendix~\ref{aa:activeLayers}.

\section{Discussion}\label{s:discussion}

The comparison of the experimental and simulation results is not straightforward, and it is necessary to consider many factors that may affect the interpretation of the results, including the assignment of the species detected in the experiment. We discuss these factors in what follows.

\subsection{Uncertainties in the
OCS, CS, and SO formation pathways}
\label{ss:uncertainties}

The fact that the simulation predicts substantial amounts of OCS, CS, and SO in stark contradiction with the experimental results may be due to several effects. Perhaps the most obvious potential reason for the discrepancy is uncertainties pertaining to the reactions considered in the chemical model, the majority of which are considered to be barrierless (see Appendix~\ref{aa:chemicalNetwork}).

\begin{figure}
\centering
        \includegraphics[width=.65\columnwidth]{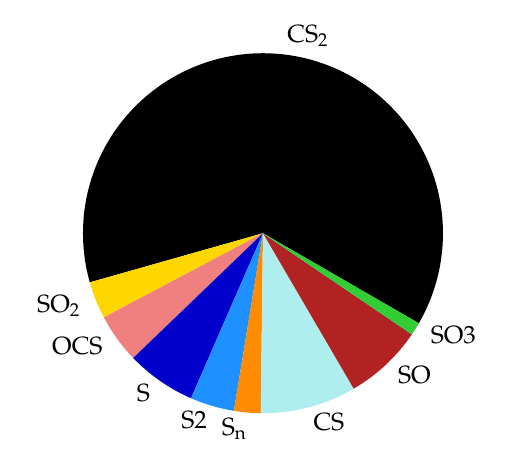}
    \caption{As the left-hand pie chart in Fig.\,\ref{fig:nondiff100ML}, but showing the results of the test model where a small activation barrier of 1.5\,K is introduced for the CO + S reaction.}  
    \label{fig:1.5Kbarrier}
\end{figure}

Let us first examine OCS. The experiments of \citet{MartinDomenech24} employed oxygen and carbon isotopes so that the formation pathways of the species formed during the experiments could be quantified. They concluded that $\sim$75\% of OCS formation was due to CS + O, and $\sim$25\% due to CO + S. Our fiducial simulation predicts the reverse, with 55\% of OCS formation due to CO + S, 34\% due to CS + O, and the remaining 11\% due to other minor formation channels. It is not surprising that the results differ in this regard considering that the simulation predicts a significant amount of CS and SO, and very likely the CS/SO and S/O ratios differ from those in the experiment as well, although the information required to confirm this cannot be recovered from the experiment. As a very simple test to evaluate the effect of the reactions in question, we introduced an activation barrier to the CO + S reaction in order to favor OCS production via CS + O instead. We found that an extremely low barrier of 1.5\,K is enough to make the model produce OCS via CS + O in a roughly 3:1 ratio to the production rate via CO + S. The number densities predicted by this test model are shown in Fig.\,\ref{fig:1.5Kbarrier}. Evidently, the atomic S freed up this way does not go on to produce increased amounts of $\rm S_2$ or ${\rm S}_n$, but mostly remains in atomic form. Apart from OCS and S, the results are very similar to those shown in Fig.\,\ref{fig:nondiff100ML}.

For SO, the main formation pathways, as given by the simulation output, are $\rm O + S \longrightarrow SO$ and $\rm OCS + O \longrightarrow CO + SO$, while its destruction occurs mainly through oxidation reactions into SO$_2$ via $\rm SO + O \longrightarrow SO_2$ and $\rm SO + SO \longrightarrow SO_2 + S$. All of these reactions are assumed to be barrierless. It is difficult to promote the destruction of SO and hence the formation of $\rm SO_2$ in the simulation to get the results closer to the experiments. The main source of atomic oxygen is the dissociation of $\rm CO_2$, which is already (slightly) overestimated by the model (Fig.\,\ref{fig:CS2_CO2_decay}), meaning that the number density of atomic O is not the main issue. Then, presumably, the $\rm SO + O  \longrightarrow SO_2$ reaction would need to be made more efficient for the available SO to be converted into $\rm SO_2$. There is however no natural way of accomplishing this with the present simulation setup\footnote{The implementation of nondiffusive chemistry takes into account, for example, the fact that reaction products can form in an excited state which allows to surmount activation barriers more easily in an immediate follow-on reaction. But since most reactions in the network are barrierless, this property of the model does not induce more reactivity than is expected when excitation is not considered.}, short of artificially increasing the reaction rate (see also Sect.\,\ref{ss:excitation} for discussion on the diffusion of atomic S). Interestingly, further oxidation of SO$_2$ into SO$_3$ occurs to a greater extent in the simulation compared to the experiment (Fig. \ref{fig:nondiff100ML}). 
In contrast, previous experimental works \citep[][]{pilling15,bonfim17} suggested that the formation of SO$_3$ in irradiated SO$_2$-bearing ices is expected to be hindered in mixtures compared to pure SO$_2$ samples.

Similarly to the case of SO, there is no natural way within the constraints of the simulation parameters to remove the ``excess'' CS compared to the experimental results (the amount of CS is very difficult to establish experimentally; we discuss this in more detail in Sect.\,\ref{sec:as_so_exp}). According to the simulation data, 97\% of CS formation is due to the dissociation of $\rm CS_2$, while 56\% of its destruction is due to OCS formation (23\% goes to reforming $\rm CS_2$ through $\rm CS + S$, and all other processes are minor). Therefore, reducing the amount of CS would require a reduction in the $\rm CS_2$ dissocation rate, but then the amount of $\rm CS_2$ predicted by the simulation would no longer match the experimental result.

The simulations described above do not take into account the so-called reaction-diffusion competition process, where reaction rates can be limited by the reactant diffusing away from the binding site before a barrier-mediated reaction with a neighboring molecule has a chance to occur (see Eq.\,5 in \citealt{Jin20}). The competition can be optionally enabled in {\sl pyRate}, and we have run a repeat simulation of the fiducial model to see if the competition would make a significant different to our results. Although the introduction of the competition does influence reaction rates, the results of these tests are qualitatively similar to the fiducial model results, which again reflects the fact that most of the reactions in the network are barrierless.

We conclude from the comparison of the fiducial simulation results to the experimental findings, and based on the additional simulation tests, that the discrepancies between the two are very likely due to incomplete knowledge of the chemical reactions at play. It may be that some of the barrierless reactions in fact have an activation barrier, and/or
some significant chemical pathways might be missing from the model. 
A good example of the importance of the latter is the $\rm OCS + O \longrightarrow CO + SO$ reaction \citep{Maity13}, which was added for the present work (see also Appendix~\ref{aa:chemicalNetwork}), and 
turned out to be
the major OCS destruction pathway in our model. Missing reactions may also contribute to the excess abundance of CS. 
Regarding barriers, nondiffusive chemistry is very efficient and, indeed, even small activation barriers can play a large role. 
Dedicated investigations into the possibility of activation barriers for the most important reactions, and of the possibility of previously unconsidered reactions, should be carried out in the future.

\subsection{Role of excitation in the formation of ${\rm S}_n$}\label{ss:excitation}

Rate-equation based methods of simulating grain-surface chemistry do not conventionally take excitation effects into account (though the nondiffusive chemistry model to some extent does, as pointed out above). \citet{Shingledecker18} presented a formulation of radiolysis-driven grain-surface chemistry (termed ``radiation chemistry'') which takes into account enhanced reactivity due to the excitation of the species. The method was applied in \citet{Shingledecker20a} to investigate the abundances of sulfur allotropes. They found a significant (many orders of magnitude) boost to sulfur allotrope production when the radiolytic processes were accounted for. This functionality is at present not included in {\sl pyRate}, and hence we could not test the effect of radiation chemistry on our results -- specifically to investigate the potential boost to sulfur allotrope formation. However, we note that the models presented in \citet{Shingledecker20a} predict a substantial amount of OCS formation, and so it is unclear whether and to what extent radiation chemistry could in fact help to favor allotropes over OCS in the present case.

Regarding the experiments in \citet{MartinDomenech24}, it is possible that the atomic S originating in $\rm CS_2$ dissociation formed in an excited state, and that the excited S atom could have transiently moved around the surface at enhanced efficiency until relaxation back to the ground state occured. More mobility for the S atoms should in principle promote the formation of the sulfur allotropes as well as the formation of $\rm SO_2$ through $\rm O + SO$ (preceded by $\rm O + S$). 
As a test, we ran a simulation where the diffusion energy of atomic S was decreased to half of its fiducial value\footnote{We assume a binding energy of atomic S of 1100\,K, and a diffusion-to-binding energy ratio of 0.55, giving a fiducial diffusion energy of 605\,K.} -- that is, overestimating the effect of transiently enhanced diffusion by permanently increasing the mobility of atomic S. This led to a decrease of the number density of atomic S, and an associated slight increase in ${\rm S}_n$; the changes compared to the fiducial model are however on the per cent level only. The number densities of other species remained almost unchanged, and hence the test model failed to increase the formation of $\rm SO_2$ or decrease the production of OCS, in particular. Therefore, while we cannot exclude excitation effects as a possible cause for the differences between the experimental and simulation results, their impact is probably lower than that of the uncertainties in the chemical network (as described in Sect.\,\ref{ss:uncertainties}).

\subsection{Detectability
of CS and SO in the experiment and other experimental uncertainties}\label{sec:as_so_exp}

While the previous Sections discuss potential shortcomings in the model that could be responsible for the discrepancies between the simulation and the experimental results,  
this Section revisits the analysis of experiment 5 in \citet{MartinDomenech24} to explore the nondetection of CS and SO in the irradiated ice mixture, and the amount of sulfur estimated to be locked in allotropes.  

\begin{figure}
    \centering
    \includegraphics[width=0.6\linewidth]{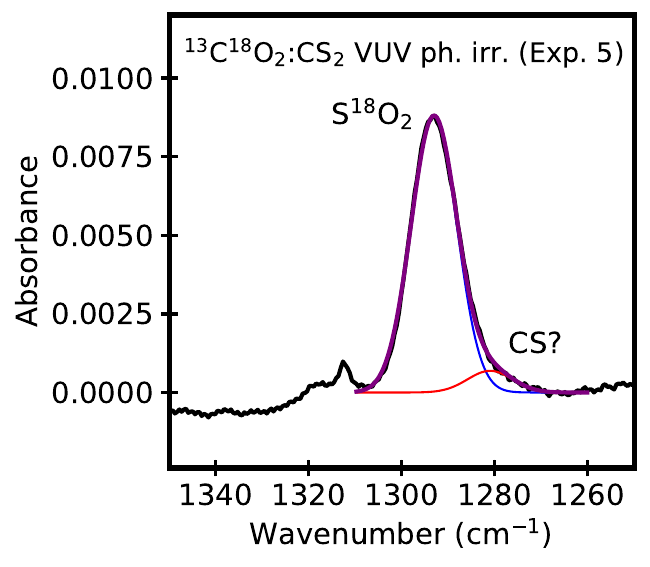}
    \caption{IR spectra in the 1350$-$1250 cm$^{-1}$ region of a VUV photon irradiated $^{13}$C$^{18}$O$_2$:CS$_2$ ice sample \citep[experiment 5 in][]{MartinDomenech24}. A two-Gaussian fit (purple) was applied to the observed band, with the blue line assigned to S$^{18}$O$_2$, and the red line potentially corresponding to CS.}
    \label{fig:cs_exp5}
\end{figure}

\begin{figure}
    \centering
    \includegraphics[width=1\linewidth]{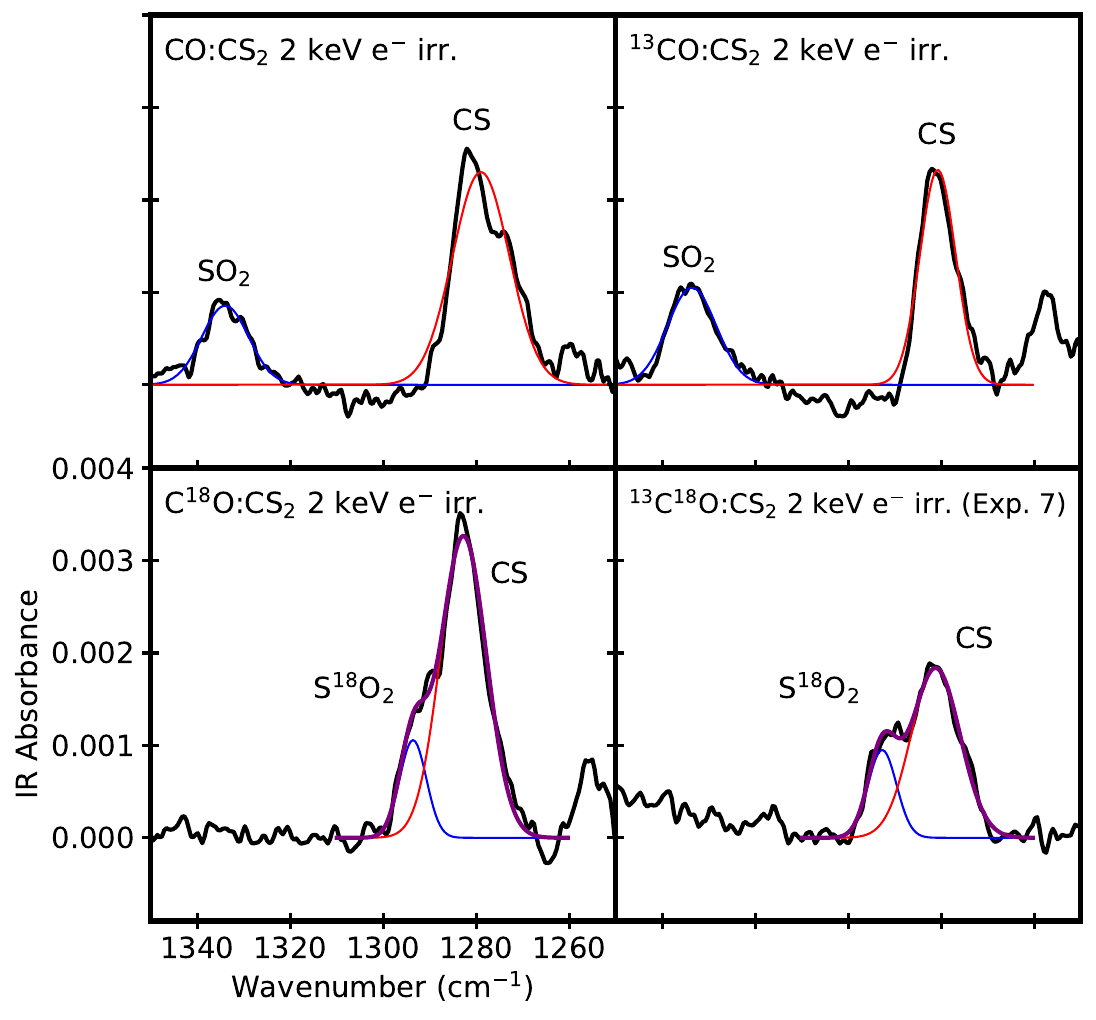}
    \caption{IR spectra in the 1350$-$1250 cm$^{-1}$ region of 2 keV electron irradiated CO:CS$_2$ (top left), $^{13}$CO:CS$_2$ (top right), C$^{18}$O:CS$_2$ (bottom left), and $^{13}$C$^{18}$O:CS$_2$ (bottom right) ice samples (in black). Red lines represent Gaussian fits of the IR feature assigned to CS, while blue lines correspond to the S$^{18}$O$_2$ feature.}
    \label{fig:cs_co_13co}
\end{figure}

\begin{table*}[]
    \centering
    \begin{tabular}{cccc}
       Molecule & Transition & Band position [$\rm cm^{-1}$] & Reference\\
       \hline
       \hline
        SO$_2$ & $\nu_3$ & 1335 & \cite{Maity13}\\
        S$^{18}$O$_2$ & $\nu_3$ & 1290 &  \cite{Maity13}\\
        CS & $\nu_1$ & 1280$-$1270 & \cite{bahou00}\\
        SO$_2$ & $\nu_1$ & 1148 & \cite{Maity13}\\
        SO & $\nu_1$ & 1137 &  \cite{lo04}\\
        S$^{18}$O$_2$ & $\nu_1$ & 1096 & \cite{Maity13}\\
        S$^{18}$O & $\nu_1$ & 1092 &  \cite{lo04}\\
        \hline
    \end{tabular}
    \caption{IR band positions of S-bearing molecules present in irradiated CO$_2$:CS$_2$ ices.}
    \label{tab:irbands}
\end{table*}

According to \citet{bahou00}, CS molecules present an IR feature in the 1270$-$1280 cm$^{-1}$ range, depending on the ice environment (see Table \ref{tab:irbands} for this and other S-bearing molecule band positions). 
Unfortunately, in experiment 5 of \citet{MartinDomenech24} (where the $^{13}$C$^{18}$O$_2$ isotopolog was used) this position overlapped with the S$^{18}$O$_2$ IR feature corresponding to the antisymmetric stretch vibration mode \citep[1290 cm$^{-1}$,][]{Maity13}. 
In this experiment, the IR band detected at $\sim$1293 cm$^{-1}$ (Fig. \ref{fig:cs_exp5}) was assigned solely to S$^{18}$O$_2$. 
The possibility that the CS feature could be hindered by the S$^{18}$O$_2$ band was mentioned in \citet{MartinDomenech24}, but it was not explored further. 
To evaluate the detectability of the CS IR feature under these circumstances, we analyzed additional irradiation experiments of CO:CS$_2$ ice mixtures using different CO isotopologs, including experiment 7 in \citet{MartinDomenech24}. 
In these experiments, formation of both SO$_2$ and CS molecules was expected, with the former proceeding to a lower extent compared to CO$_2$:CS$_2$ ices \citep{MartinDomenech24}.
The corresponding IR spectra in the 1350$-$1250 cm$^{-1}$ region are shown in Fig. \ref{fig:cs_co_13co}.
In experiments with $^{16}$O isotopologs the SO$_2$ IR band appeared at $\sim$1335 cm$^{-1}$ \citep{Maity13}, enabling the detection of both the SO$_2$ and CS IR features (top panels of Fig. \ref{fig:cs_co_13co}). 
On the other hand, when $^{18}$O isotopologs were used, the S$^{18}$O$_2$ IR feature shifted from $\sim$1335 cm$^{-1}$ to $\sim$1293 cm$^{-1}$, overlapping with the CS feature. 
Nevertheless, two distinct peaks were observed, and we were able to 
disentangle both contributions with a two-Gaussian fit
(bottom panels of Fig. \ref{fig:cs_co_13co}).
Following the theoretical prediction that a significant fraction of the initial sulfur in irradiated CO$_2$:CS$_2$ ices could be contained in CS molecules,
we applied the same two-Gaussian fit
to the spectrum shown in Fig. \ref{fig:cs_exp5}, in order to evaluate the presence of a hindered CS IR feature. 
In this case, the fit was degenerated because only one peak was detected. Therefore, we limited the position of the CS feature to the 1275$-$1281 cm$^{-1}$ range in order to get a similar fit to those obtained when the two peaks were observed. 
Interestingly, the detected IR band previously assigned to S$^{18}$O$_2$ presented a small excess in the red wing compared to a Gaussian profile. This is compatible with the presence of a weaker CS feature with an integrated absorbance of $\sim$0.0085 cm$^{-1}$. 
The corresponding CS column density would depend on the value of the IR band strength, which is unfortunately not known. 

\begin{figure}
    \centering
    \includegraphics[width=0.6\linewidth]{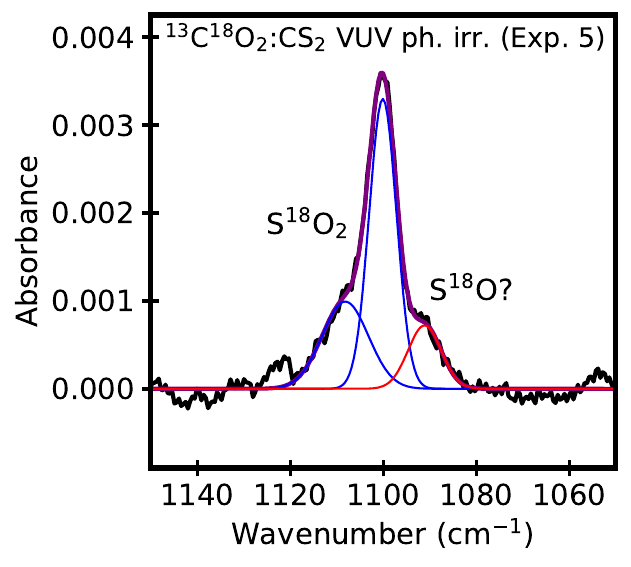}
    \caption{IR spectra in the 1150$-$1050 cm$^{-1}$ region of a VUV photon  irradiated $^{13}$C$^{18}$O$_2$:CS$_2$ ice sample \citep[experiment 5 in][]{MartinDomenech24}. A three-Gaussian fit (purple) was applied to the observed band, with the blue lines assigned to S$^{18}$O$_2$, and the red line potentially corresponding to S$^{18}$O.}
    \label{fig:so_exp5}
\end{figure}

A similar scenario was found when exploring the presence of SO in experiment 5. 
According to the band positions reported in \citet{lo04} and \citet{Maity13}, the SO IR feature (located at 1137 cm$^{-1}$ for the regular isotopolog, or 1092 cm$^{-1}$ for S$^{18}$O) overlaps with the SO$_2$ IR band corresponding to the symmetric stretch vibration mode (peaking at $\sim$1150 cm$^{-1}$ for the regular isotopolog, or $\sim$1100 cm$^{-1}$ for S$^{18}$O$_2$). 
Similar to the S$^{18}$O$_2$ IR band shown in Fig. \ref{fig:cs_exp5}, the band assigned to the symmetric stretch also presented an excess in both wings compared to a Gaussian profile (Fig. \ref{fig:so_exp5}).  
In this case, we performed a three-Gaussian fitting of the observed band, limiting the width parameter of all Gaussians to a maximum value of~5. 
The two Gaussians peaking at $\sim$1108 cm$^{-1}$ and $\sim$1100 cm$^{-1}$ were assigned to S$^{18}$O$_2$, while the third Gaussian located at $\sim$1091 cm$^{-1}$ could correspond to S$^{18}$O, with an integrated absorbance of $\sim$0.0065 cm$^{-1}$.
As for CS, the corresponding S$^{18}$O abundance depends on the unknown IR band strength. Therefore, a more careful analysis of the IR spectrum of experiment 5 in \citet{MartinDomenech24} reveals that the experimental results are indeed compatible with the presence of some CS and SO molecules.

Another discrepancy between the model and the experiment is the amount of sulfur potentially locked in allotropes. In this regard, the presence of CS and SO would reshape the distribution of sulfur at the end of the experiment shown in the right panel of Fig. \ref{fig:nondiff100ML}, and thus the amount of undetected sulfur assigned to allotropes. According to Fig. \ref{fig:nondiff100ML}, the detected SO$_2$, OCS, C$_3$S$_2$, and SO$_3$ in experiment 5 represented, approximately, 10\%, 0.5\%, 0.5\%, and 0.3\% of the initial sulfur (respectively), while $\sim$61\% of the initial sulfur remained in CS$_2$ molecules. 
The missing $\sim$27\% of sulfur was proposed to be locked in sulfur allotropes. However, including the CS and SO abundances in the sulfur balance would lower this amount.
Unfortunately, the band strengths of their IR features are not known, as mentioned above.
This introduces a large uncertainty, and we can only speculate on their column densities. 
In the case of CS, 
if we assume the same band strength as CS$_2$ \citep[1.1 $\times$ 10$^{-16}$ cm molecule$^{-1}$,][]{Taillard2025}, the column density at the end of experiment 5 would be $\sim$0.2 ML. If the band strength was an order of magnitude lower\footnote{For reference, the CO band strength is a factor of $\sim$7 lower than the CO$_2$ band strength \citep{gerakines95}.} (close to the value estimated in \citealt{basalgete26}), the CS abundance could be up to 2 ML. 
This would represent between 0.25\% and 2.5\% of the initial sulfur in experiment 5. 
In the case of SO (S$^{18}$O in the experiment), the column density would be 0.35$-$3.5 ML, representing 0.5$-$5\% of the initial sulfur, depending on whether we assume the same band strength as for the SO$_2$ antisymmetric stretch \citep[4.2 $\times$ 10$^{-17}$ cm molecule$^{-1}$,][]{yarnall22} or an order of magnitude lower, respectively. 
In the scenario with higher CS and SO abundances (assuming the lower end of the speculated band strengths), they could account for up to $\sim$8\% of the initial sulfur (a somewhat lower abundance than that predicted by the model).  
In addition, we note that the band strength used in \citet{MartinDomenech24} to calculate the C$_3$S$_2$ abundance was also roughly approximated. An order of magnitude lower value, for example, would lead to C$_3$S$_2$ representing up to 5\% of the initial sulfur. 
This (along with the higher end of the estimated CS and SO abundances) would cut in half the amount of undetected sulfur at the end of the experiment. 
Another source of uncertainty in the sulfur balance discussed in \citet{MartinDomenech24} is the 20\% uncertainty assumed for the reported IR band strengths. This would mostly affect the amount of sulfur contained in CS$_2$ and SO$_2$ (the most abundant detected S-bearing molecules at the end of the experiment).
In the least favorable scenario (the initial sulfur was overestimated by 20\% and the produced SO$_2$ was underestimated by 20\%), the amount of undetected sulfur that could be contained in allotropes would be further decreased to $\sim$6\%, which is closer to the theoretical results shown in the middle panel of Fig. \ref{fig:nondiff100ML}.
Therefore, even though we cannot confirm the presence of sulfur allotropes (or any other undetectable sulfur sink) in the irradiated ice, we consider that it is likely according to the experimental and theoretical results, and to previous experiments with other S-bearing ices \citep[see, e.g.,][]{hector24}.

\section{Astrophysical implications}\label{s:implications}

A significant fraction of sulfur in dense and cold interstellar regions is expected to be locked in icy dust mantles \citep[e.g.,][]{laas19}. However, only two sulfur-bearing ice species, OCS and tentatively SO$_2$, have been detected so far, and their combined abundances account for less than 5\% of the total sulfur budget. Other compounds, such as sulfur allotropes, polysulfanes, and the ammonium salt NH$_4$SH, have been proposed as major sulfur reservoirs in the ice \citep[see, e.g.,][]{cazaux22,hector24,Slavicinska2025}. The detection of sulfur allotropes and polysulfanes remains extremely challenging even with state-of-the-art instrumentation \citep[see, e.g.,][]{Taillard2025}. In the case of NH$_4$SH, \citet{Slavicinska2025} derived an upper limit to the amount of sulfur that could be stored in this salt based on JWST detections of NH$_4^+$; even in the unrealistic scenario in which all detected $\rm NH_4^+$ originated from NH$_4$SH, this reservoir would account for only $\sim$20\% of the sulfur column toward the source.

Thus, the identity of the main sulfur reservoir in interstellar ices remains an open question. In the absence of direct detections, chemical models and laboratory experiments provide the most promising avenue to infer the composition of icy mantles. Ideally, chemical models should be validated by comparing their predictions with a representative sample of observed species in space. However, the abundances of gas-phase sulfur species depend not only on formation and destruction pathways but also on adsorption and desorption processes linking the gas and solid phases. Large uncertainties in nonthermal and thermal desorption mechanisms hamper any robust inference of ice composition from gas-phase data alone. Moreover, the limited observational constraints on sulfur-bearing ices prevent a direct comparison between chemical networks and the true ice composition \citep{Taillard2025}. Under these conditions, simulating laboratory experiments using chemical models appears as the most reliable strategy to test current networks and to explore how different physical and chemical parameters shape the sulfur budget in the ice.

In this work, the {\sl pyRate} code has been used to simulate the formation of OCS and SO$_2$ (along with other S-bearing species) in a CO$_2$:CS$_2$ ice, reproducing experiment 5 from \citet{MartinDomenech24}. This experiment irradiates a CO$_2$:CS$_2$ ice with UV photons at a temperature of 10 K. During the experiment, a set of heavier sulfur molecules, mainly SO$_2$ and OCS, but also C$_3$S$_2$ and SO$_3$ are formed. In addition, a fraction of sulfur remain undetected 
at the end of the experiment.
This missing sulfur was attributed to the formation of sulfur chains (S$_n$).
The chemical network used by {\sl pyRate} includes all the formation and destruction mechanisms needed to form these compounds, including the formation of allotropes as described by \cite{Shingledecker19}, with the sole exception of the minor species $\rm C_3S_2$ whose formation has not been simulated here. In a first trial, we used diffusive chemistry to reproduce the chemical processes taking place during the experiment. Our simulations showed that diffusive chemistry alone cannot form any of the detected molecules, except for the direct products of CS$_2$ photodissociation, under the experimental conditions. This is not unexpected since the diffusion of heavy species on the grain surfaces is very limited at 10\,K. The comparison of the simulations and experiments indicates that nondiffusive chemistry is instead required to reproduce the formation of heavy sulfur-bearing species under cold core conditions.

\begin{figure}
    \centering
    \includegraphics[width=0.49\textwidth]{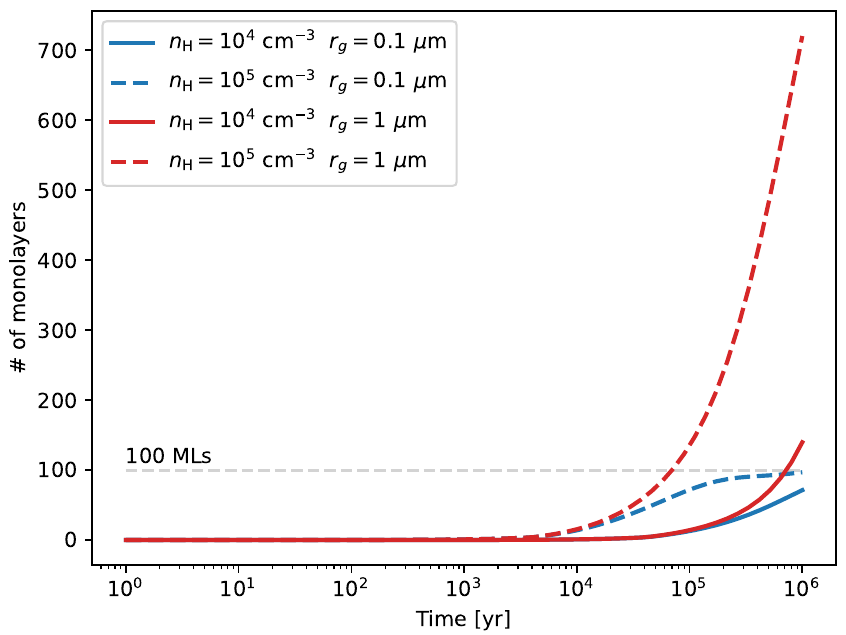}
    \caption{Number of ice MLs as a function of time for two densities, $n_{\rm H}=10^4$ cm$^{-3}$ (solid line) and $10^{5}$ cm$^{-3}$ (dashed line), and two average grain radii, $r_{g}=0.1\ \mu m$ (blue) and $1\ \mu m$ (red). The 100 MLs level is shown as a horizontal dashed grey line.}
    \label{fig:nMLs}
\end{figure}

The simulations adopt the total of 530 ice monolayers deposited in the experiment, but the predicted abundances are sensitive to the number of monolayers that are affected by the incoming UV radiation, which could not be constrained in the analysis of the experiments. However, as we show in Appendix~\ref{aa:activeLayers}, the simulation results are qualitatively similar whether the amount of active layers is 100 or 300. Overall the best agreement with the experimental results is obtained for 100 active layers which is why we have chosen it as our fiducial model. This contrasts with chemical models typically assuming 1-4 active surface MLs (e.g., \citealt{Vasyunin13a,Taquet14,Sipila15a,ruaud16}). We tested whether this value is a good representation of the expected number of icy layers in typical astrochemical settings. For instance, we took the model developed for Barnard 1b by \citet{NavarroAlmaida2025} and its predicted chemical abundances through time to compute the total number of ice monolayers. We considered four cases: two values for the gas density, $n_{\rm H}=10^4,\ 10^{5}$ cm$^{-3}$, and two average grain radii $r_{g}=0.1,\ 1$ $\mu m$. These densities and grain sizes reflect typical conditions found in translucent and dense gas toward molecular clouds, where grain sizes increase by one order of magnitude \citep[see, e.g.,][]{Steinacker2015}. Recent JWST observations support this grain size enhancement toward molecular clouds \citep{Dartois2024}. If the dust-to-gas mass ratio is kept constant and grains are spherical, the total grain surface area and the number of adsorption sites decrease with larger grains \citep{NavarroAlmaida2024}, leading to the increasing number of MLs that is seen in Fig.\,\ref{fig:nMLs}, where we show the number of ice MLs deposited as a function of time for the four cases. In molecular cloud conditions we therefore expect the presence of at least 100\,MLs. Even with smaller grains at higher densities, 100\,MLs is a good estimate for the final number of icy layers. Considering that for thinner active surfaces the simulations predict little reactivity, a 100 ML thickness for the active surface is an adequate representation of the UV-induced action in the ice.

Finally, the experiments allow an assessment of how well current chemical networks reproduce the formation of long sulfur chains. The models seem to underestimate the production of sulfur allotropes, 
while overpredicting the abundances of simple sulfur species such as CS and SO. 
Even when the experimental uncertainties are taken into account, the results 
suggest that the chemical network governing the formation of these molecules is incomplete and requires substantial revision.

\section{Summary and conclusions}\label{s:conclusions}
We used the {\sl pyRate} astrochemical model to simulate the VUV irradiation of a CO$_2$:CS$_2$ ice mixture at 10 K under laboratory conditions.  This study represents the first effort to model the chemistry of a multicomponent ice analog using a rate-equation–based code. Our main results can be summarized as follows:

\begin{itemize}

\item We compare the ability of diffusive and nondiffusive chemistry to reproduce the UV irradiation experiment. Our results demonstrate that nondiffusive chemistry is required to explain the formation of S-bearing species at temperatures as low as 10 K.

\item The comparison also shows that the number of chemically active layers is a key parameter in simulations adopting nondiffusive chemistry. The best agreement between the experiment and the {\sl pyRate} simulations is obtained when considering a thickness of 100 monolayers (ML) for the active surface. Increasing the number of ML beyond this value does not significantly improve the results, even though the experiment used 530\,ML. Since the simulations predict very little reactivity in thinner ices, our work provides an experimental-based estimate of the UV penetration depth in interstellar ices.

\item The chemical simulations successfully reproduce the set of molecular species detected in the experiment, although discrepancies remain in their relative abundances. While OCS, CS, and SO are the dominant S-bearing products predicted by the model, SO$_2$ is the most abundant species observed experimentally. In addition, a significant fraction of sulfur remains undetected in the experiment, possibly locked in sulfur allotropes.

\item These discrepancies likely arise from a combination of incomplete chemical networks (missing reactions or poorly constrained reaction barriers) and uncertainties in the experimental analysis. In particular, some fraction of sulfur may be locked in CS and SO in the experiment but remain hidden due to the overlap of their infrared bands with those of more abundant species.

\item From an astrochemical perspective, our experiments and simulations indicate that properly modeling sulfur chemistry in cold regions (T $\sim$10 K), such as starless cores, requires 
the inclusion of nondiffusive surface chemistry and an extended number of chemically active layers.

\end{itemize}
Overall, this work highlights the complementary role of laboratory experiments and astrochemical modeling in constraining sulfur chemistry in interstellar ices and advancing our understanding of the missing sulfur problem.

\begin{acknowledgements}
The authors are grateful to the referee, Prof. Sergio Pilling, for constructive comments that helped to improve the manuscript. O.S. and W.R. thank the Max Planck Society for financial support. This project has also received funding from “la Caixa” Foundation under agreement LCF/BQ/PI22/11910030, from the Spanish Ministry of Science and Innovation through project PID2023-151513NB-C21, and from  the European Research Council (ERC) under the European Union’s Horizon Europe research and innovation programme ERC-AdG-2022 (SUL4LIFE GA No.\ 101096293). Funded by the European Union. Views and opinions expressed are however those of the author(s) only and do not necessarily reflect those of the European Union or the European Research Council Executive Agency. Neither the European Union nor the granting authority can be held responsible for them.
\end{acknowledgements}

\bibliographystyle{aa}
\bibliography{aa59941-26}

\begin{appendix}
\onecolumn

\section{Effect of the number of chemically active ice monolayers}\label{aa:activeLayers}

\begin{figure*}[h]
\centering
\includegraphics[width=1.0\columnwidth]{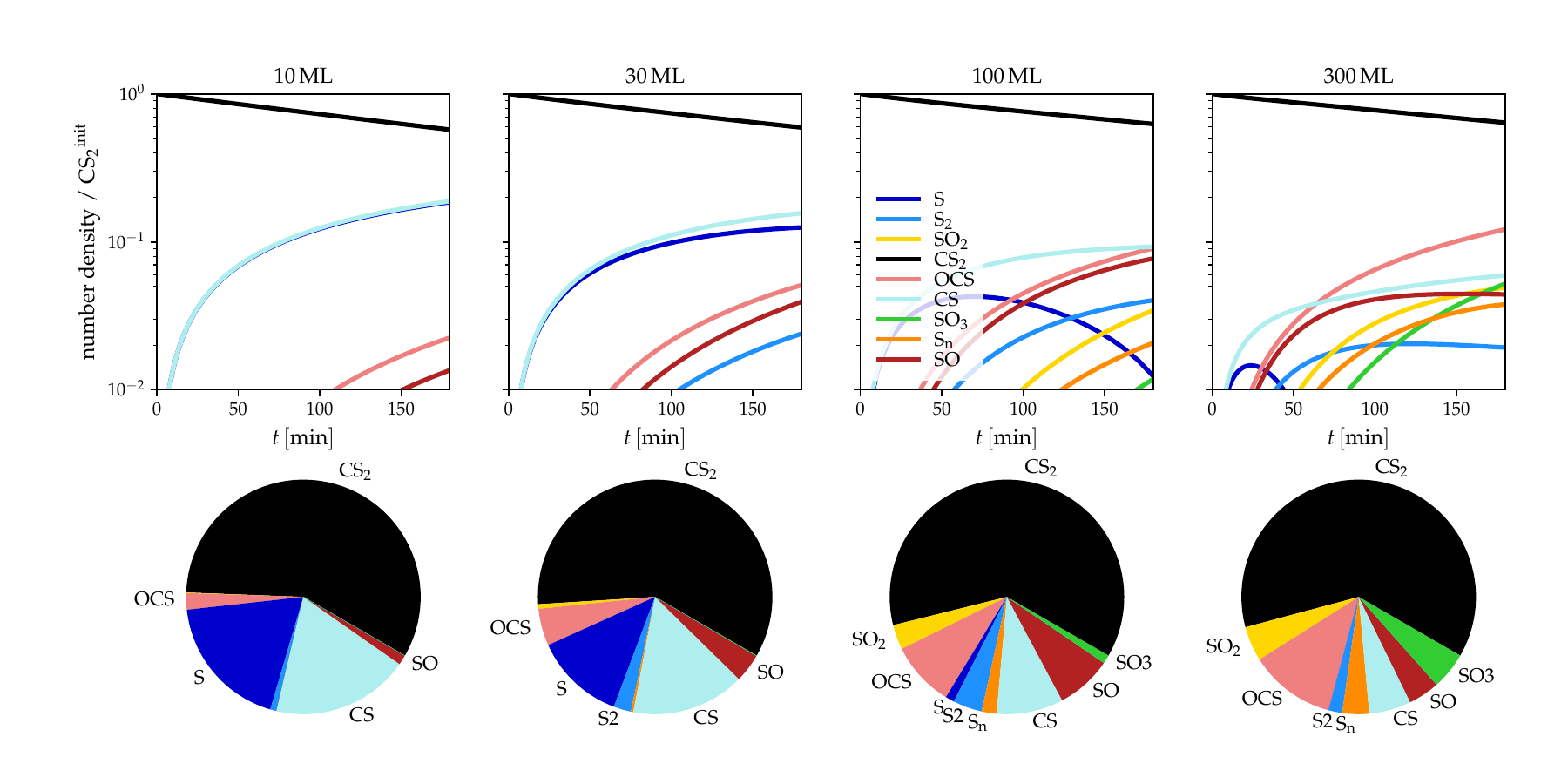}
    \caption{As Fig.\,\ref{fig:nondiff100ML} in the main text, but showing results for different monolayer thicknesses of the active surface (as indicated on top of each panel). For clarity, only those species with a relative abundance of over 1\,\% are labeled on the pie charts. The fiducial 100\,ML model is also included for reference.}
        \label{fig:MLvariation}
\end{figure*}

To quantify the effect of the number of chemically active monolayers on the simulation results, we have run a number of simulations where the thickness (in MLs) of the active surface is varied between 10 and 300\,ML. Figure~\ref{fig:MLvariation} shows the results of these for four different active surface thicknesses.

Firstly, it is evident that little chemical evolution occurs in the timescale of the experiment if the active surface is thin. Such behavior is expected given the (relatively) short duration of the experiment; with few reactants available, chemical evolution will proceed slowly and not much will occur over a short time interval. When the thickness of the active layer is increased, the conversion of atomic S increases in efficiency, and species such as OCS and SO start to form in abundance.

Secondly, the results of the 100 and 300\,ML active surface models are qualitatively similar. Increasing the surface thickness from the fiducial value of 100 ML to 300\,ML boosts the production of $\rm SO_3$ at the cost of SO; however, the amounts of CS and SO are still substantial and in particular there is even more OCS produced than in the fiducial simulation. Allotrope production is boosted slightly but remains well below the level deduced from the experiments. We conclude from Fig.\,\ref{fig:MLvariation} that the discrepancies between the experiment and the simulations are very likely not due to the uncertainty in the relative amount of the ice that the incoming UV radiation acts on, but -- as we discuss in the main text -- rather to the uncertainties in the chemical network.

\section{Chemical network}\label{aa:chemicalNetwork}

Table~\ref{tab:reactionNetwork} shows the reactions included in the chemical simulations. The base network is taken from \citet{Semenov10}; for the present work, all reactions that contain elements other than carbon, oxygen, or sulfur have been removed. Most reactions involving sulfur have been taken from \cite{laas19}, with two reactions not included there adopted from \citet{Maity13}. The rate coefficients of some photoreactions have been updated based on the corresponding values in kida.uva.2014 \citep{Wakelam15} or on the KIDA website when newer values are available. We assume the rate coefficients of ice photodissociation reactions to be equal to those in the gas phase. The effect of this assumption on our results is examined in Appendix~\ref{aa:photorates}.

For the photoreactions, the parameters $\alpha$ and $\beta$ are respectively the prefactor and exponential attenuation coefficients (the rate coefficient is calculated with the formula $k = \alpha \, \exp (-\beta \, A_{\rm V})$). For the other reactions, the two parameters represent respectively the branching ratio and activation energy in~K.

\onecolumn
\begin{longtable}{cccccccccccc}
\caption{Chemical network considered in this work.}\\
\hline\hline
\centering

& & Chemical reaction & & & & & $\alpha$ & $\beta$  & Reference \\ \hline
\endfirsthead
\caption{continued.}\\
\hline\hline
& & Chemical reaction & & & & & $\alpha$ & $\beta$  & Reference \\ \hline
\endhead
\endlastfoot

$\rm CO$ & $\rm PHOTON$ & $\longrightarrow$ & $\rm C$ & $\rm O$ & $\rm $ & $\rm $ & 2.60e-10 & 3.53e+00 & KIDA\footnote{Data incorporated either from the KIDA 2014 public release \citep{Wakelam15}, or from the KIDA website when newer values are available. The adopted values correspond to gas-phase reactions; we assume that the same ones are valid in ice as well.}\\ 
$\rm CO_{2}$ & $\rm PHOTON$ & $\longrightarrow$ & $\rm CO$ & $\rm O$ & $\rm $ & $\rm $ & 8.90e-10 & 3.00e+00 & KIDA\\ 
$\rm CS$ & $\rm PHOTON$ & $\longrightarrow$ & $\rm C$ & $\rm S$ & $\rm $ & $\rm $ & 9.80e-10 & 2.43e+00 & KIDA\\ 
$\rm C_{2}$ & $\rm PHOTON$ & $\longrightarrow$ & $\rm C$ & $\rm C$ & $\rm $ & $\rm $ & 2.40e-10 & 2.57e+00 & KIDA\\ 
$\rm CCO$ & $\rm PHOTON$ & $\longrightarrow$ & $\rm CO$ & $\rm C$ & $\rm $ & $\rm $ & 5.00e-10 & 1.70e+00 & \citet{Semenov10}\\ 
$\rm CCO$ & $\rm PHOTON$ & $\longrightarrow$ & $\rm C_{2}$ & $\rm O$ & $\rm $ & $\rm $ & 5.00e-10 & 1.70e+00 & \citet{Semenov10}\\ 
$\rm C_{3}$ & $\rm PHOTON$ & $\longrightarrow$ & $\rm C_{2}$ & $\rm C$ & $\rm $ & $\rm $ & 5.00e-09 & 2.07e+00 & KIDA\\ 
$\rm C_{3}O$ & $\rm PHOTON$ & $\longrightarrow$ & $\rm C_{2}$ & $\rm CO$ & $\rm $ & $\rm $ & 7.00e-09 & 1.58e+00 & KIDA\\ 
$\rm C_{3}S$ & $\rm PHOTON$ & $\longrightarrow$ & $\rm C_{2}$ & $\rm CS$ & $\rm $ & $\rm $ & 1.00e-10 & 2.00e+00 & \citet{Semenov10}\\ 
$\rm C_{4}$ & $\rm PHOTON$ & $\longrightarrow$ & $\rm C_{2}$ & $\rm C_{2}$ & $\rm $ & $\rm $ & 1.28e-09 & 2.30e+00 & KIDA\\ 
$\rm C_{4}$ & $\rm PHOTON$ & $\longrightarrow$ & $\rm C_{3}$ & $\rm C$ & $\rm $ & $\rm $ & 7.23e-09 & 2.30e+00 & KIDA\\ 
$\rm C_{4}S$ & $\rm PHOTON$ & $\longrightarrow$ & $\rm C_{3}$ & $\rm CS$ & $\rm $ & $\rm $ & 1.00e-10 & 2.00e+00 & \citet{Semenov10}\\ 
$\rm C_{5}$ & $\rm PHOTON$ & $\longrightarrow$ & $\rm C_{3}$ & $\rm C_{2}$ & $\rm $ & $\rm $ & 8.50e-12 & 2.30e+00 & KIDA\\ 
$\rm C_{6}$ & $\rm PHOTON$ & $\longrightarrow$ & $\rm C_{5}$ & $\rm C$ & $\rm $ & $\rm $ & 1.00e-10 & 1.70e+00 & KIDA\\ 
$\rm C_{6}$ & $\rm PHOTON$ & $\longrightarrow$ & $\rm C_{4}$ & $\rm C_{2}$ & $\rm $ & $\rm $ & 1.00e-10 & 1.70e+00 & KIDA\\ 
$\rm C_{7}$ & $\rm PHOTON$ & $\longrightarrow$ & $\rm C_{5}$ & $\rm C_{2}$ & $\rm $ & $\rm $ & 2.00e-10 & 1.70e+00 & KIDA\\ 
$\rm C_{7}$ & $\rm PHOTON$ & $\longrightarrow$ & $\rm C_{4}$ & $\rm C_{3}$ & $\rm $ & $\rm $ & 8.00e-10 & 1.70e+00 & KIDA\\ 
$\rm C_{8}$ & $\rm PHOTON$ & $\longrightarrow$ & $\rm C_{7}$ & $\rm C$ & $\rm $ & $\rm $ & 5.00e-11 & 1.70e+00 & KIDA\\ 
$\rm C_{9}$ & $\rm PHOTON$ & $\longrightarrow$ & $\rm C_{7}$ & $\rm C_{2}$ & $\rm $ & $\rm $ & 5.00e-11 & 1.70e+00 & KIDA\\ 
$\rm C_{9}$ & $\rm PHOTON$ & $\longrightarrow$ & $\rm C_{6}$ & $\rm C_{3}$ & $\rm $ & $\rm $ & 6.50e-10 & 1.70e+00 & KIDA\\ 
$\rm C_{9}$ & $\rm PHOTON$ & $\longrightarrow$ & $\rm C_{5}$ & $\rm C_{4}$ & $\rm $ & $\rm $ & 3.00e-10 & 1.70e+00 & KIDA\\ 
$\rm C_{10}$ & $\rm PHOTON$ & $\longrightarrow$ & $\rm C_{7}$ & $\rm C_{3}$ & $\rm $ & $\rm $ & 7.50e-10 & 1.70e+00 & KIDA\\ 
$\rm C_{10}$ & $\rm PHOTON$ & $\longrightarrow$ & $\rm C_{5}$ & $\rm C_{5}$ & $\rm $ & $\rm $ & 2.50e-10 & 1.70e+00 & KIDA\\ 
$\rm O_{2}$ & $\rm PHOTON$ & $\longrightarrow$ & $\rm O$ & $\rm O$ & $\rm $ & $\rm $ & 7.90e-10 & 2.13e+00 & KIDA\\ 
$\rm S_{2}$ & $\rm PHOTON$ & $\longrightarrow$ & $\rm S$ & $\rm S$ & $\rm $ & $\rm $ & 3.30e-10 & 1.40e+00 & KIDA\\ 
$\rm SO$ & $\rm PHOTON$ & $\longrightarrow$ & $\rm O$ & $\rm S$ & $\rm $ & $\rm $ & 4.20e-09 & 2.37e+00 & KIDA\\ 
$\rm SO_{2}$ & $\rm PHOTON$ & $\longrightarrow$ & $\rm SO$ & $\rm O$ & $\rm $ & $\rm $ & 1.90e-09 & 2.38e+00 & KIDA\\ 
$\rm OCS$ & $\rm PHOTON$ & $\longrightarrow$ & $\rm CO$ & $\rm S$ & $\rm $ & $\rm $ & 4.67e-09 & 2.46e+00 & KIDA\\ 
$\rm C$ & $\rm C$ & $\longrightarrow$ & $\rm C_{2}$ & $\rm $ & $\rm $ & $\rm $ & 1.00e+00 & 0.00e+00 & \citet{Semenov10}\\ 
$\rm C$ & $\rm C_{2}$ & $\longrightarrow$ & $\rm C_{3}$ & $\rm $ & $\rm $ & $\rm $ & 1.00e+00 & 0.00e+00 & \citet{Semenov10}\\ 
$\rm C$ & $\rm CCO$ & $\longrightarrow$ & $\rm C_{3}O$ & $\rm $ & $\rm $ & $\rm $ & 1.00e+00 & 0.00e+00 & \citet{Semenov10}\\ 
$\rm C$ & $\rm C_{3}$ & $\longrightarrow$ & $\rm C_{4}$ & $\rm $ & $\rm $ & $\rm $ & 1.00e+00 & 0.00e+00 & \citet{Semenov10}\\ 
$\rm C$ & $\rm C_{4}$ & $\longrightarrow$ & $\rm C_{5}$ & $\rm $ & $\rm $ & $\rm $ & 1.00e+00 & 0.00e+00 & \citet{Semenov10}\\ 
$\rm C$ & $\rm C_{5}$ & $\longrightarrow$ & $\rm C_{6}$ & $\rm $ & $\rm $ & $\rm $ & 1.00e+00 & 0.00e+00 & \citet{Semenov10}\\ 
$\rm C$ & $\rm C_{6}$ & $\longrightarrow$ & $\rm C_{7}$ & $\rm $ & $\rm $ & $\rm $ & 1.00e+00 & 0.00e+00 & \citet{Semenov10}\\ 
$\rm C$ & $\rm C_{7}$ & $\longrightarrow$ & $\rm C_{8}$ & $\rm $ & $\rm $ & $\rm $ & 1.00e+00 & 0.00e+00 & \citet{Semenov10}\\ 
$\rm C$ & $\rm C_{8}$ & $\longrightarrow$ & $\rm C_{9}$ & $\rm $ & $\rm $ & $\rm $ & 1.00e+00 & 0.00e+00 & \citet{Semenov10}\\ 
$\rm C$ & $\rm C_{9}$ & $\longrightarrow$ & $\rm C_{10}$ & $\rm $ & $\rm $ & $\rm $ & 1.00e+00 & 0.00e+00 & \citet{Semenov10}\\ 
$\rm C$ & $\rm O$ & $\longrightarrow$ & $\rm CO$ & $\rm $ & $\rm $ & $\rm $ & 1.00e+00 & 0.00e+00 & \citet{Semenov10}\\ 
$\rm C$ & $\rm O_{2}$ & $\longrightarrow$ & $\rm CO$ & $\rm O$ & $\rm $ & $\rm $ & 1.00e+00 & 0.00e+00 & \citet{Semenov10}\\ 
$\rm C$ & $\rm S$ & $\longrightarrow$ & $\rm CS$ & $\rm $ & $\rm $ & $\rm $ & 1.00e+00 & 0.00e+00 & \citet{Semenov10}\\ 
$\rm C$ & $\rm SO$ & $\longrightarrow$ & $\rm CO$ & $\rm S$ & $\rm $ & $\rm $ & 1.00e+00 & 0.00e+00 & \citet{Semenov10}\\ 
$\rm O$ & $\rm C_{2}$ & $\longrightarrow$ & $\rm CCO$ & $\rm $ & $\rm $ & $\rm $ & 1.00e+00 & 0.00e+00 & \citet{Semenov10}\\ 
$\rm O$ & $\rm C_{3}$ & $\longrightarrow$ & $\rm C_{3}O$ & $\rm $ & $\rm $ & $\rm $ & 1.00e+00 & 0.00e+00 & \citet{Semenov10}\\ 
$\rm O$ & $\rm CO$ & $\longrightarrow$ & $\rm CO_{2}$ & $\rm $ & $\rm $ & $\rm $ & 1.00e+00 & 1.00e+03 & \citet{Semenov10}\\ 
$\rm O$ & $\rm CS$ & $\longrightarrow$ & $\rm OCS$ & $\rm $ & $\rm $ & $\rm $ & 1.00e+00 & 0.00e+00 & \citet{Semenov10}\\ 
$\rm O$ & $\rm O$ & $\longrightarrow$ & $\rm O_{2}$ & $\rm $ & $\rm $ & $\rm $ & 1.00e+00 & 0.00e+00 & \citet{Semenov10}\\ 
$\rm O$ & $\rm S$ & $\longrightarrow$ & $\rm SO$ & $\rm $ & $\rm $ & $\rm $ & 1.00e+00 & 0.00e+00 & \citet{Semenov10}\\ 
$\rm O$ & $\rm SO$ & $\longrightarrow$ & $\rm SO_{2}$ & $\rm $ & $\rm $ & $\rm $ & 1.00e+00 & 0.00e+00 & \citet{Semenov10}\\ 
$\rm S$ & $\rm CO$ & $\longrightarrow$ & $\rm OCS$ & $\rm $ & $\rm $ & $\rm $ & 1.00e+00 & 0.00e+00 & \citet{Semenov10}\\ 
$\rm C$ & $\rm C_{2}S$ & $\longrightarrow$ & $\rm C_{2}$ & $\rm CS$ & $\rm $ & $\rm $ & 1.00e+00 & 0.00e+00 & \citet{laas19}\\ 
$\rm C$ & $\rm C_{3}S$ & $\longrightarrow$ & $\rm C_{3}$ & $\rm CS$ & $\rm $ & $\rm $ & 1.00e+00 & 0.00e+00 & \citet{laas19}\\ 
$\rm C$ & $\rm C_{4}S$ & $\longrightarrow$ & $\rm C_{4}$ & $\rm CS$ & $\rm $ & $\rm $ & 1.00e+00 & 0.00e+00 & \citet{laas19}\\ 
$\rm C$ & $\rm OCS$ & $\longrightarrow$ & $\rm CO$ & $\rm CS$ & $\rm $ & $\rm $ & 1.00e+00 & 0.00e+00 & \citet{laas19}\\
$\rm C$ & $\rm S_{2}$ & $\longrightarrow$ & $\rm CS_{2}$ & $\rm $ & $\rm $ & $\rm $ & 1.00e+00 & 0.00e+00 & \citet{laas19}\\ 
$\rm C$ & $\rm SO$ & $\longrightarrow$ & $\rm CS$ & $\rm O$ & $\rm $ & $\rm $ & 1.00e+00 & 0.00e+00 & \citet{laas19}\\ 
$\rm C$ & $\rm SO_{2}$ & $\longrightarrow$ & $\rm CO$ & $\rm SO$ & $\rm $ & $\rm $ & 1.00e+00 & 0.00e+00 & \citet{laas19}\\ 
$\rm C_{2}$ & $\rm S$ & $\longrightarrow$ & $\rm C$ & $\rm CS$ & $\rm $ & $\rm $ & 1.00e+00 & 0.00e+00 & \citet{laas19}\\ \hline
$\rm C_{2}O$ & $\rm S$ & $\longrightarrow$ & $\rm CO$ & $\rm CS$ & $\rm $ & $\rm $ & 1.00e+00 & 0.00e+00 & \citet{laas19}\\ 
$\rm C_{2}S$ & $\rm O$ & $\longrightarrow$ & $\rm CO$ & $\rm CS$ & $\rm $ & $\rm $ & 1.00e+00 & 0.00e+00 & \citet{laas19}\\ 
$\rm C_{2}S$ & $\rm S$ & $\longrightarrow$ & $\rm CS$ & $\rm CS$ & $\rm $ & $\rm $ & 1.00e+00 & 0.00e+00 & \citet{laas19}\\ 
$\rm C_{3}S$ & $\rm O$ & $\longrightarrow$ & $\rm C_{2}S$ & $\rm CO$ & $\rm $ & $\rm $ & 1.00e+00 & 2.31e+02 & \citet{laas19}\\ 
$\rm C_{4}$ & $\rm S$ & $\longrightarrow$ & $\rm C_{3}$ & $\rm CS$ & $\rm $ & $\rm $ & 1.00e+00 & 0.00e+00 & \citet{laas19}\\ 
$\rm C_{4}S$ & $\rm O$ & $\longrightarrow$ & $\rm C_{3}S$ & $\rm CO$ & $\rm $ & $\rm $ & 5.00e-01 & 0.00e+00 & \citet{laas19}\\ 
$\rm C_{4}S$ & $\rm O$ & $\longrightarrow$ & $\rm C_{3}O$ & $\rm CS$ & $\rm $ & $\rm $ & 5.00e-01 & 0.00e+00 & \citet{laas19}\\ 
$\rm C_{4}S$ & $\rm S$ & $\longrightarrow$ & $\rm C_{3}S$ & $\rm CS$ & $\rm $ & $\rm $ & 1.00e+00 & 0.00e+00 & \citet{laas19}\\ 
$\rm C_{6}$ & $\rm S$ & $\longrightarrow$ & $\rm C_{5}$ & $\rm CS$ & $\rm $ & $\rm $ & 1.00e+00 & 0.00e+00 & \citet{laas19}\\ 
$\rm CS$ & $\rm O_{2}$ & $\longrightarrow$ & $\rm CO$ & $\rm SO$ & $\rm $ & $\rm $ & 5.00e-01 & 0.00e+00 & \citet{laas19}\\ 
$\rm CS$ & $\rm O_{2}$ & $\longrightarrow$ & $\rm O$ & $\rm OCS$ & $\rm $ & $\rm $ & 5.00e-01 & 0.00e+00 & \citet{laas19}\\ 
$\rm CS$ & $\rm S$ & $\longrightarrow$ & $\rm CS_{2}$ & $\rm $ & $\rm $ & $\rm $ & 1.00e+00 & 0.00e+00 & \citet{laas19}\\ 
$\rm CS_{2}$ & $\rm O$ & $\longrightarrow$ & $\rm CS$ & $\rm SO$ & $\rm $ & $\rm $ & 3.33e-01 & 8.20e+02 & \citet{laas19}\\ 
$\rm CS_{2}$ & $\rm O$ & $\longrightarrow$ & $\rm OCS$ & $\rm S$ & $\rm $ & $\rm $ & 3.33e-01 & 8.20e+02 & \citet{laas19}\\ 
$\rm CS_{2}$ & $\rm O$ & $\longrightarrow$ & $\rm CO$ & $\rm S_{2}$ & $\rm $ & $\rm $ & 3.33e-01 & 8.20e+02 & \citet{laas19}\\ 
$\rm O$ & $\rm S_{2}$ & $\longrightarrow$ & $\rm S$ & $\rm SO$ & $\rm $ & $\rm $ & 1.00e+00 & 0.00e+00 & \citet{laas19}\\ 
$\rm O_{2}$ & $\rm S$ & $\longrightarrow$ & $\rm O$ & $\rm SO$ & $\rm $ & $\rm $ & 1.00e+00 & 0.00e+00 & \citet{laas19}\\ 
$\rm O_{2}$ & $\rm SO$ & $\longrightarrow$ & $\rm O$ & $\rm SO_{2}$ & $\rm $ & $\rm $ & 1.00e+00 & 2.30e+03 & \citet{laas19}\\ 
$\rm O_{3}$ & $\rm S$ & $\longrightarrow$ & $\rm O_{2}$ & $\rm SO$ & $\rm $ & $\rm $ & 1.00e+00 & 0.00e+00 & \citet{laas19}\\ 
$\rm O_{3}$ & $\rm SO$ & $\longrightarrow$ & $\rm O_{2}$ & $\rm SO_{2}$ & $\rm $ & $\rm $ & 1.00e+00 & 1.20e+03 & \citet{laas19}\\ 
$\rm OCS$ & $\rm S$ & $\longrightarrow$ & $\rm CO$ & $\rm S_{2}$ & $\rm $ & $\rm $ & 1.00e+00 & 0.00e+00 & \citet{laas19}\\ 
$\rm S$ & $\rm S$ & $\longrightarrow$ & $\rm S_{2}$ & $\rm $ & $\rm $ & $\rm $ & 1.00e+00 & 0.00e+00 & \citet{laas19}\\ 
$\rm S$ & $\rm S_{2}$ & $\longrightarrow$ & $\rm S_{3}$ & $\rm $ & $\rm $ & $\rm $ & 1.00e+00 & 0.00e+00 & \citet{laas19}\\ 
$\rm S$ & $\rm S_{3}$ & $\longrightarrow$ & $\rm S_{4}$ & $\rm $ & $\rm $ & $\rm $ & 1.00e+00 & 0.00e+00 & \citet{laas19}\\ 
$\rm S$ & $\rm S_{4}$ & $\longrightarrow$ & $\rm S_{5}$ & $\rm $ & $\rm $ & $\rm $ & 1.00e+00 & 0.00e+00 & \citet{laas19}\\ 
$\rm S$ & $\rm S_{5}$ & $\longrightarrow$ & $\rm S_{6}$ & $\rm $ & $\rm $ & $\rm $ & 1.00e+00 & 0.00e+00 & \citet{laas19}\\ 
$\rm S_{2}$ & $\rm S_{2}$ & $\longrightarrow$ & $\rm S_{4}$ & $\rm $ & $\rm $ & $\rm $ & 1.00e+00 & 0.00e+00 & \citet{laas19}\\ 
$\rm S_{2}$ & $\rm S_{3}$ & $\longrightarrow$ & $\rm S_{5}$ & $\rm $ & $\rm $ & $\rm $ & 1.00e+00 & 0.00e+00 & \citet{laas19}\\ 
$\rm S_{2}$ & $\rm S_{4}$ & $\longrightarrow$ & $\rm S_{6}$ & $\rm $ & $\rm $ & $\rm $ & 1.00e+00 & 0.00e+00 & \citet{laas19}\\ 
$\rm S_{2}$ & $\rm S_{5}$ & $\longrightarrow$ & $\rm S_{7}$ & $\rm $ & $\rm $ & $\rm $ & 1.00e+00 & 0.00e+00 & \citet{laas19}\\ 
$\rm S_{3}$ & $\rm S_{3}$ & $\longrightarrow$ & $\rm S_{6}$ & $\rm $ & $\rm $ & $\rm $ & 1.00e+00 & 0.00e+00 & \citet{laas19}\\ 
$\rm S_{3}$ & $\rm S_{4}$ & $\longrightarrow$ & $\rm S_{7}$ & $\rm $ & $\rm $ & $\rm $ & 1.00e+00 & 0.00e+00 & \citet{laas19}\\ 
$\rm S_{3}$ & $\rm S_{5}$ & $\longrightarrow$ & $\rm S_{8}$ & $\rm $ & $\rm $ & $\rm $ & 1.00e+00 & 0.00e+00 & \citet{laas19}\\ 
$\rm S_{4}$ & $\rm S_{4}$ & $\longrightarrow$ & $\rm S_{8}$ & $\rm $ & $\rm $ & $\rm $ & 1.00e+00 & 0.00e+00 & \citet{laas19}\\ 
$\rm SO$ & $\rm SO$ & $\longrightarrow$ & $\rm S$ & $\rm SO_{2}$ & $\rm $ & $\rm $ & 1.00e+00 & 0.00e+00 & \citet{laas19}\\ 
$\rm CS_{2}$ & $\rm PHOTON$ & $\longrightarrow$ & $\rm CS$ & $\rm S$ & $\rm $ & $\rm $ & 8.80e-09 & 2.50e+00 & \citet{laas19}\\ 
$\rm S_{3}$ & $\rm PHOTON$ & $\longrightarrow$ & $\rm S$ & $\rm S$ & $\rm S$ & $\rm $ & 1.00e-10 & 0.00e+00 & \citet{laas19}\\ 
$\rm S_{4}$ & $\rm PHOTON$ & $\longrightarrow$ & $\rm S$ & $\rm S_{3}$ & $\rm $ & $\rm $ & 6.00e-11 & 0.00e+00 & \citet{laas19}\\ 
$\rm S_{5}$ & $\rm PHOTON$ & $\longrightarrow$ & $\rm S$ & $\rm S_{4}$ & $\rm $ & $\rm $ & 3.00e-11 & 0.00e+00 & \citet{laas19}\\ 
$\rm S_{6}$ & $\rm PHOTON$ & $\longrightarrow$ & $\rm S_{3}$ & $\rm S_{3}$ & $\rm $ & $\rm $ & 1.50e-11 & 0.00e+00 & \citet{laas19}\\ 
$\rm S_{7}$ & $\rm PHOTON$ & $\longrightarrow$ & $\rm S$ & $\rm S_{6}$ & $\rm $ & $\rm $ & 2.50e-12 & 0.00e+00 & \citet{laas19}\\ 
$\rm S_{7}$ & $\rm PHOTON$ & $\longrightarrow$ & $\rm S_{3}$ & $\rm S_{4}$ & $\rm $ & $\rm $ & 2.50e-12 & 0.00e+00 & \citet{laas19}\\ 
$\rm S_{8}$ & $\rm PHOTON$ & $\longrightarrow$ & $\rm S_{4}$ & $\rm S_{4}$ & $\rm $ & $\rm $ & 1.00e-12 & 0.00e+00 & \citet{laas19}\\ 
$\rm S_{8}$ & $\rm PHOTON$ & $\longrightarrow$ & $\rm S_{3}$ & $\rm S_{5}$ & $\rm $ & $\rm $ & 1.00e-12 & 0.00e+00 & \citet{laas19}\\ 
$\rm OCS$ & $\rm O$ & $\longrightarrow$ & $\rm CO$ & $\rm SO$ & $\rm $ & $\rm $ & 1.00e+00 & 0.00e+00 & \citet{Maity13}\\ 
$\rm SO_{2}$ & $\rm O$ & $\longrightarrow$ & $\rm SO_{3}$ & $\rm $ & $\rm $ & $\rm $ & 1.00e+00 & 0.00e+00 & \citet{Maity13}\\ 
$\rm SO_{3}$ & $\rm PHOTON$ & $\longrightarrow$ & $\rm SO_{2}$ & $\rm O$ & $\rm $ & $\rm $ & 1.90e-09 & 2.38e+00 & This work\footnote{Set equal to the rate coefficient of $\rm SO_2 + PHOTON$.}\\ 
\hline

\label{tab:reactionNetwork}
\end{longtable}

\section{Uncertainties in the rate coefficients of photoreactions}\label{aa:photorates}

\begin{figure*}
\centering
\includegraphics[width=.8\columnwidth]{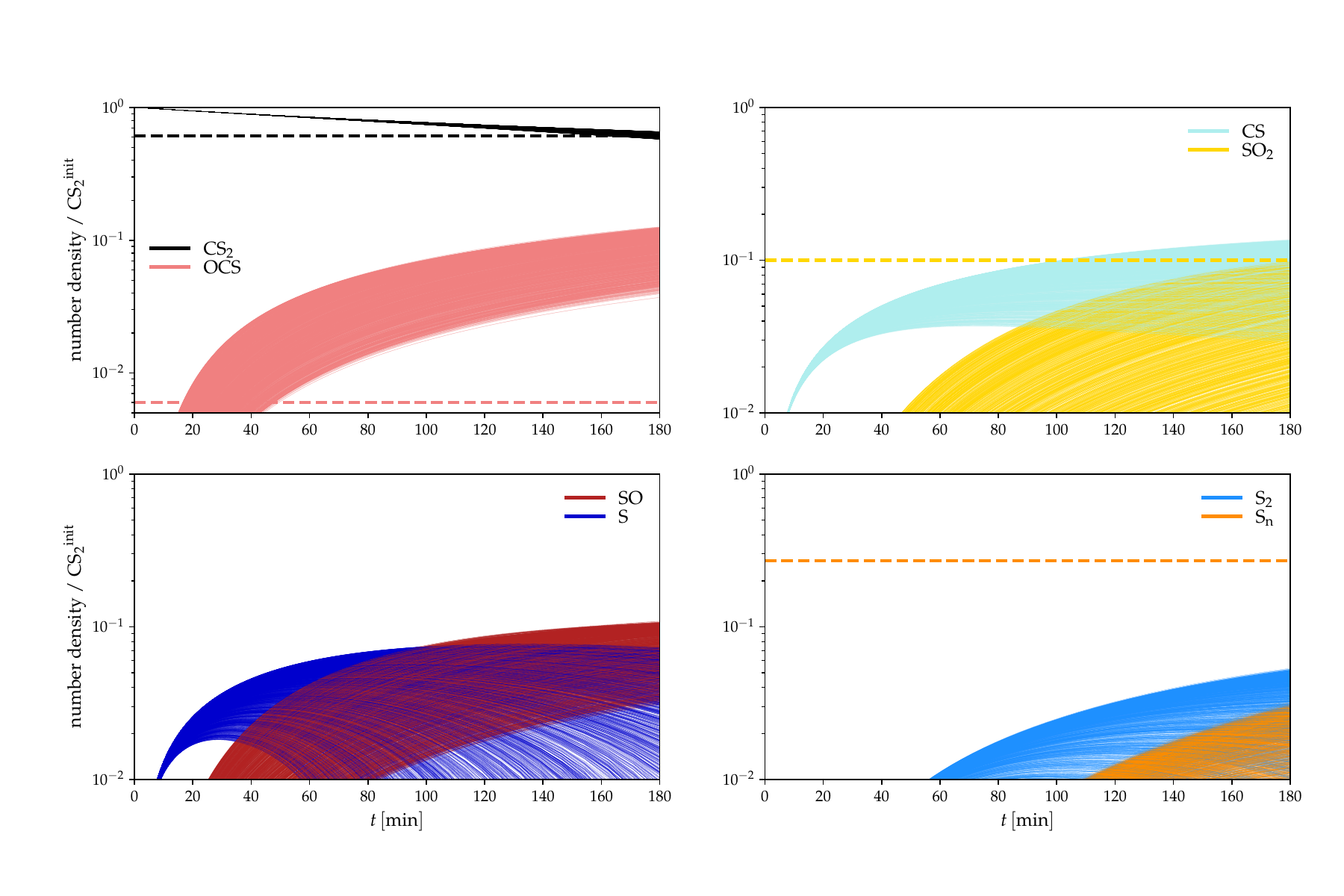}
     \caption{Time-evolution of the number densities of selected sulfur carriers, normalized to the initial number density of $\rm CS_2$, in the set of 2000 simulations where the fiducial nondiffusive model (Fig.\,\ref{fig:nondiff100ML}) is modified by introducing random variation in the photodissociation rate coefficients. Horizontal lines represent the corresponding experimental values (cf. Fig.\,\ref{fig:nondiff100ML}).}
        \label{fig:randomizedSimulations}
\end{figure*}

\begin{figure*}
\centering
\begin{picture}(600,150)(0,0)
\put(0,0){
\begin{picture}(0,0) 
\includegraphics[width=0.75\columnwidth]{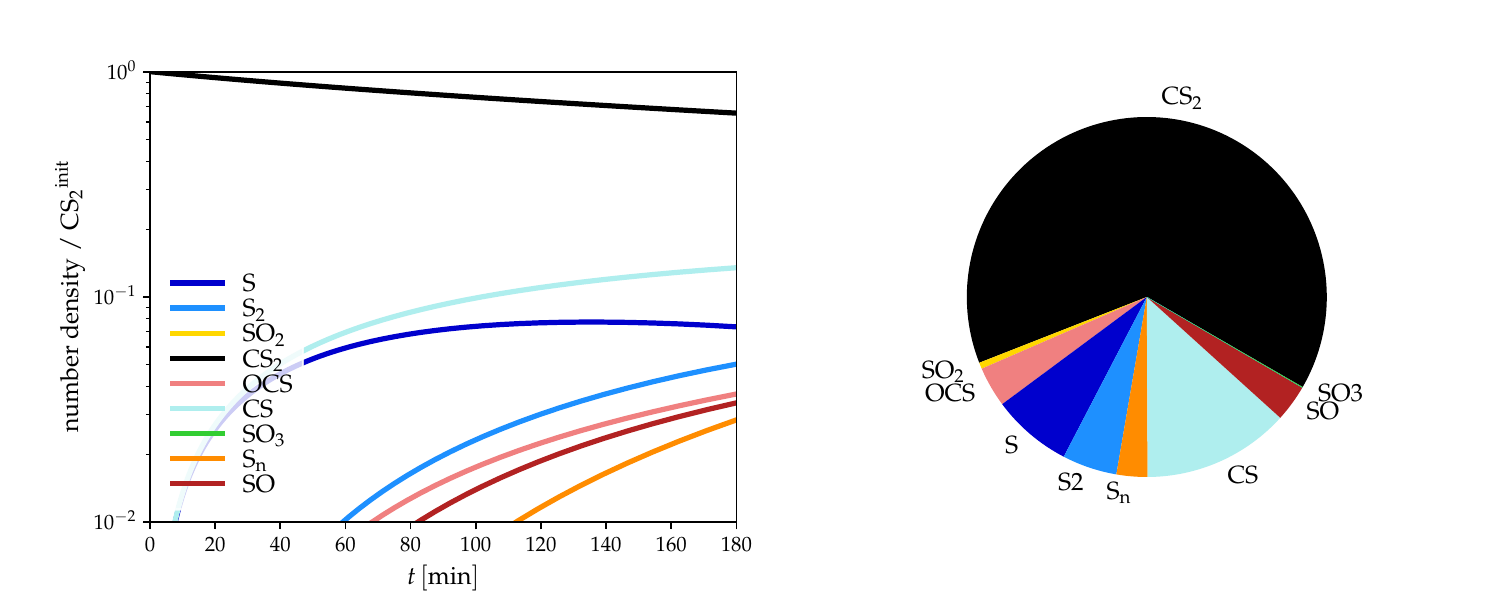}
\end{picture}}
\put(370,22){
\begin{picture}(0,0) 
        \includegraphics[width=0.25\columnwidth]{figures/13c18o2_cs2_s_chemistry_Exp5_Olli.pdf}
\end{picture}}
\end{picture}
    \caption{As Fig.\,\ref{fig:nondiff100ML} in the main text, but showing the results of the simulation which minimizes the OCS fraction at 180 min (Fig.\,\ref{fig:randomizedSimulations}).}
        \label{fig:minimizedOCS}
\end{figure*}

As noted in Appendix~\ref{aa:chemicalNetwork}, the chemical network employed here assumes that the rate coefficients of photoreactions in the ice are equal to those of the gas-phase counterpart reactions. This is a common assumption in simulations when no dedicated rate coefficient information is available for ice species, and in fact \citet{laas19}, whose data we adopt presently, made the same assumption. Nevertheless, it is prudent to examine the sensitivity of our results to uncertainties in the photodissocation rate coefficients, given that reactivity is in the experiments induced by VUV photons. To this end, adapting the methodology laid out in \citet{JimenezRedondo24}, we ran a set of 2000 simulations where the rate coefficients of all photoreactions, except for $\rm CS_2$ dissociation, were varied randomly by a factor of 3 each way from their values in the fiducial model (i.e., the nondiffusive model depicted in Fig.\,\ref{fig:nondiff100ML}). $\rm CS_2$ dissocation was not modified because it initiates the S chemistry, and with a modified rate coefficient the model would no longer match the experimental $\rm CS_2$ dissociation curve (Fig.\,\ref{fig:CS2_CO2_decay}).

Figure~\ref{fig:randomizedSimulations} shows the results of the simulations, with normalized number density curves from all 2000 simulations overlaid. Evidently, by varying the photodissociation rate coefficients it is possible to obtain solutions where the normalized OCS number density is approximately 4\% at 180 min of simulation time (as opposed to $\sim$9\% in the fiducial model). Greatly elevated $\rm SO_2$ fractions are also possible. However, SO and CS always have non-negligible fractions, and the $S_n$ fraction never rises above $\sim$3\%. We note that the small variation in the $\rm CS_2$ abundance is a back-effect due to the variable reactivity among the 2000 simulations.

Figure~\ref{fig:minimizedOCS} shows the results of the simulation where the OCS fraction at 180 min is minimized. Although the OCS fraction is substantially reduced, the predicted amount of CS is even larger than in the fiducial simulation, and $\rm SO_2$ is produced only in small amounts.

The results of these tests lead us to conclude that the assumption of equality between gas-phase and ice photodissociation rates is not the underlying reason behind the discrepancies between the simulation results and the experiments. At the same time, it is clear that uncertainties in the photodissociation rates do have a marked influence on the simulation results and that, in the future, attention should be paid to constraining photodissociation rates in ices.

\end{appendix}
\end{document}